\documentclass[10pt,journal,compsoc]{IEEEtran}
\usepackage{amsthm}
\usepackage{cite}
\usepackage{amsmath,amssymb,amsfonts}
\usepackage{graphicx}
\usepackage{fancyhdr}
\usepackage{textcomp}
\usepackage{xcolor}
\usepackage{algorithmicx,algpseudocode}
\usepackage{algpseudocode,algorithm}
\usepackage{booktabs}
\usepackage{kotex}

\algrenewcommand\algorithmicrequire{\textbf{Precondition:}}
\algrenewcommand\algorithmicensure{\textbf{Postcondition:}}
\theoremstyle{definition}

\newtheorem{proposition}{Proposition}

\usepackage{caption}
\usepackage{subcaption}

\usepackage[normalem]{ulem}
\begin{document}
\title{Delay-Sensitive and Power-Efficient Quality Control of Dynamic Video Streaming using Adaptive Super-Resolution}

\author{Minseok Choi,~\IEEEmembership{Member, IEEE},
    Won Joon Yun,~\IEEEmembership{Student Member, IEEE}, 
    and 
    Joongheon Kim,~\IEEEmembership{Senior Member, IEEE}
\IEEEcompsocitemizethanks{\IEEEcompsocthanksitem 
M. Choi is with the the Department of Telecommunication Engineering, Jeju National University, Jeju, Korea, e-mail: ejaqmf@jejunu.ac.kr.
\IEEEcompsocthanksitem 
W. Yun and J. Kim are with the School of Electrical Engineering, Korea University, Seoul, Korea, e-mails: \{ywjoon95,joongheon\}@korea.ac.kr.
\IEEEcompsocthanksitem 
J. Kim is a corresponding author.
}
}

\IEEEtitleabstractindextext{%
\begin{abstract}
In a decade, the adaptive quality control of video streaming and the super-resolution (SR) technique have been deeply explored. 
As edge devices improved to have exceptional processing capability than ever before, streaming users can enhance the received image quality to allow the transmitter to compress the images to save its power or pursue network efficiency.
In this sense, this paper proposes a novel dynamic video streaming algorithm that adaptively compresses video chunks at the transmitter and separately enhances the quality at the receiver using SR. 
In order to allow transmission of video chunks with different compression levels and control of the computation burden, we present the adaptive SR network which is optimized by minimizing the weighted sum of losses extracted from different layer outputs. for dynamic video streaming. 
In addition, we jointly orchestrate video delivery and resource usage, and the proposed video delivery scheme balances the tradeoff well among the average video quality, the queuing delay, buffering time, transmit power, and computation power. 
Simulation results show that the proposed scheme pursues the quality-of-services (QoS) of the video streaming better than the adaptive quality control without the cooperation of the transmitter and the receiver and the non-adaptive SR network.

\end{abstract}

}

\maketitle

\IEEEdisplaynontitleabstractindextext
\IEEEpeerreviewmaketitle

\section{Introduction}\label{sec:Introduction}
\label{sec:intro}

\IEEEPARstart{R}{ecently}, with rapidly increasing smart user devices, the content delivery network has been getting more attention for supporting the excessively large global data traffic. 
As reported in \cite{cisco2019}, tens of exabytes of global data traffic are being handled on a daily basis at present, and most of the global data traffic are dominated by online video services, e.g., video streaming, video-on-demand (VoD), live streaming and VR streaming. 
Depending on these types of applications, the quality-of-services (QoS) and the quality-of-experience (QoE) of video services, such as playback stall, latency, video quality and quality fluctuations, have been largely studied \cite{CST2015Chen}. 

According to \cite{cisco2020}, the internet-of-things (IoT) devices will account for 50\% of all global networked devices by 2023; therefore, it is expected that online video services in the IoT and/or vehicular networks will be increasingly required. 
For example, autonomous driving improves passengers' trip experiences by providing video applications with the help of vehicle-to-vehicle (V2V) communications \cite{TVT2018Guo}.
Another possible scenario is the device-to-device (D2D)-assisted wireless caching network \cite{JSAC2016Ji}.
In this case, cache-enabled devices having limited storage size and power budget can directly deliver the desired contents to streaming users; 
thus, balancing the tradeoff among communications, computations and streaming users' QoE has become {\color{blue}a} very critical issue.

As the video streaming system has launched in 
wireless networks, video applications have to be provided to users with 
limited wireless resources under the time-varying channel conditions.
In addition, heterogeneous user preferences over applications, time, and geological locations simultaneously require different quality versions of various contents. 
To deal with the above challenges, dynamic adaptive streaming over HTTP (DASH) that dynamically chooses the most appropriate bitrate has been used in wireless networks \cite{MMSys2011Stockhammer}. 
In DASH systems, a stream consists of sequential chunks and it allows video chunks to have different quality levels; therefore, bitrate adaptation schemes have been extensively studied \cite{CST2019Bentaleb}.

In parallel, with the development of edge devices having caching and computational capabilities, wireless caching and mobile edge computing (MEC) technologies have been considered as efficient methods for improving the performance of adaptive bitrate (ABR) streaming \cite{TMC2019Tran}. 
In general, video contents are encoded into multiple levels representing different bitrates and qualities; however, wireless caching helpers have 
a limited storage size so that caching 
identical content with various quality versions is inefficient. 
Therefore, MEC technologies are employed to transcode the cached contents depending on network status and/or buffer states of streaming users. 
Especially, as scalable video coding (SVC) that conveniently accumulates enhancement layers above a base layer for higher quality versions has 
gained popularity, \cite{TVT2018Zhou} joint optimization of computational tasks and video delivery has become more powerful.
On the other hand, at the user (receiver) side, edge devices can improve the quality of streams with the help of increased computing power by themselves. 
Super-resolution (SR) is 
the epitome of techniques for enhancing the video quality by using deep neural networks. 

This paper considers the situation in which the transmitter, having the capability of transcoding or compressing the desired video contents, is willing to deliver video chunks to the receiver having capability of enhancing the video quality by using SR. 
Even though the computing power and computational capability of edge devices have been developed, operation of SR is still a time- and power-consuming task.
Therefore, we focus on adaptive control of video delivery with differentiated quality requirements, transcoding at the transmitter side, and quality enhancement using SR at the receiver side. 
Video streaming over the IoT network having cache-enabled MEC entities or the vehicle-to-everything (V2X) network is a potential scenario where the transmitter 
also has limited power budget. 
Thus, our goal of the adaptive video streaming system is to pursue (i) maximizing the video quality, (ii) reducing the playback stall events, (iii) limiting power consumption of both transmitter and receiver, and (iv) stability of the video queue. 

The main contributions are as follows.

\begin{itemize}
    \item This paper employs the SR technique for dynamic video streaming, which is applicable when the transmitter and the receiver (i.e., the user) have the capability of compressing and improving the quality of video chunks, respectively. 
    In order to allow receiving video chunks and/or images with different quality levels and controlling the computational burden of SR, we employ the adaptive SR network, which is optimized by minimizing the weighted sum of losses extracted from different layer outputs.

	\item This paper proposes the adaptive quality control of video chunks depending on the time-varying network condition and both the transmitter and the receiver states. 
	We allow the transmitter to determine the video quality enhancement rate at the receiver as well as the compression rate at the transmitter, and it is beneficial to control a variety of performance metric{\color{blue}s} while improving the average quality measure.
	
	\item Joint orchestration of video delivery and resource usage of the transmitter and the receiver is proposed based on the Lyapunov optimization framework by employing the adaptive SR and allowing cooperation between the transmitter and the receiver for adaptive quality control. 
	The proposed dynamic video delivery scheme balances the tradeoff among the video quality measure, the queuing delay, the buffering time, the transmit power consumption, and the CPU usage. 
	
	\item Simulation results show that the adaptive SR network enhances the quality of input images with different compression rates and adaptively controls its computational burden, inference delay, and output quality. 
	Also, we show that the proposed dynamic video delivery balances the variety of performance metrics 
	much better than the adaptive quality control scheme without cooperation of the transmitter and the receiver, and the dynamic video streaming using the non-adaptive SR network.
\end{itemize}

The rest of the paper is organized as follows. 
The existing literature related to our work is summarized in Section \ref{sec:related_work}, and the dynamic video streaming system is described in Section \ref{sec:model}.
The adaptive control problem of video delivery and computational tasks are formulated in Section \ref{sec:joint_opt}, and the proposed adaptive control algorithm and its comparison technique are described in Section \ref{sec:schemes}{\color{blue}.}
Our simulation results are presented in Section \ref{sec:simulation}, and Section \ref{sec:conclusion} concludes this paper.

\section{Related Work}
\label{sec:related_work}

This section presents the related work of adaptive quality selection for dynamic streaming, cache/MEC-assisted ABR streaming, deep learning-based SR, and learning-based ABR streaming.

\subsection{Adaptive Quality Selection for Dynamic Streaming}

The ABR streaming has been considered 
a promising scheme for providing online video services via wireless links to users because it dynamically chooses video bitrates or quality so that streaming can efficiently adapt to time-varying environments with limited wireless resources \cite{CST2019Bentaleb}. 
Here, the fundamental issue is how to dynamically adapt the bitrate and 
how to determine which quality is appropriate 
for the current state.
The existing quality adaptation schemes are generally based on the network condition \cite{JSAC2014Li}, users' buffer states \cite{ToN2020Spiteri}, or both of them \cite{SIGCOMM2015Yin,MMSys2018He,TMCCA2019Spiteri}.

Due to the limited wireless resources and time-varying wireless channel conditions, joint optimization of quality adaptation and resource allocation has been largely studied in \cite{TCSV2015Zhao,TVT2020Khan,JSAC2018Choi}.
In \cite{TCSV2015Zhao}, scheduling and resource allocation algorithm 
that maps SVC layers to DASH layers and reduces video playback interruptions 
is presented.
Scheduling and quality selection are adaptively determined depending on both channel condition and users' buffer states in \cite{JSAC2018Choi}, and similarly, adaptive video quality selection and resource allocation method is proposed in \cite{TVT2020Khan}.

\subsection{Cache/MEC-Assisted ABR Streaming}

The ABR streaming was originated from Internet video streaming with the remote Internet server having the whole video library. 
Therefore, in the wireless streaming system, the remote Internet server fetches the desired contents to radio access networks (RAN) via wired core first network; then, the wireless RAN transmits contents to users \cite{CCNC2016Bronzino}. 
However, since fetching videos from the remote server to RAN 
can result in long latency and congestion \cite{TMM2017Ge,IWQoS2018Ren}, cache/MEC-assisted ABR streaming has been considered a promising technique for mitigating latency and congestion issues \cite{TMC2019Tran}.

The existing studies on cache/MEC-assisted ABR streaming, jointly determine the appropriate quality selection and (i) content placements \cite{ICC2017Xie,TWC2020Choi,TNSM2020Bayhan} or (ii) transcoding rate \cite{Access2017Xu,TWC2019Choi,TWC2020Choi2}. 
The cache management scheme that maximizes both the users' QoE and energy cost saving, and the probabilistic caching method for consecutive video requests are presented in \cite{ICC2017Xie} and \cite{TWC2020Choi}, respectively.
EdgeDASH, the network-side control scheme to facilitate the caching capability, is proposed for appropriate quality assignments and reducing stall events in \cite{TNSM2020Bayhan}.
Meanwhile, the authors of \cite{Access2017Xu} uses MEC to provide low-latency and ABR streaming by dynamically adapting bitrates. 
With the transcoding ability at cache-enabled nodes, link scheduling, power allocation, and delivery of individual video chunks are adaptively optimized in \cite{TWC2019Choi,TWC2020Choi2}.
Moreover, there are recent studies on dynamic streaming taking both caching and MEC into account \cite{ToN2016Pedersen,IoTJ2020Li,TMC2019Tran}.
Joint caching and transcoding policy that maximizes video capacity of the network is proposed in \cite{ToN2016Pedersen}.
The joint caching and processing framework that determines the caching method for contents with different qualities and scheduling user requests is proposed in \cite{TMC2019Tran}, and further, energy efficiency is considered as an additional performance metric in \cite{IoTJ2020Li}.

\subsection{Deep Learning-Based Super-Resolution}

The SR technique improves or recovers the quality of images or video frames, and 
modern SR techniques are majorly developed using deep learning methods. 
Among the deep learning-based SR methods, the SR convolutional neural network (SRCNN){\color{blue},}~\cite{SRCNN2015} which uses multiple convolutional layers is one of the well-known SR techniques. 
The deep learning-based SR method proposed in~\cite{VDSR2016} improves the output images by allocating image input signals into the output layer. 
The aforementioned SR algorithms evaluates results on performance numerical values only, e.g., peak-signal-to-noise-ratio (PSNR) and structural similarity index measure (SSIM).
On the other hand, generative adversarial network (GAN)-based SR methods have 
recently been proposed to pursue soft texture and smoothness of output images ~\cite{SRGAN2017,ESRGAN2018}.

While most of the existing studies focus on SR technique itself, not many have yet applied the SR to the practical video streaming and/or content delivery networks and jointly optimized the network decisions (e.g., bitrate adaptation, scheduling, and data transmission) together. 
Recently, on-device SR computation methods{\color{blue},} which enable 
the enhancement of the video quality independent at the receiver have been proposed in~\cite{HotMobile2019Hu,OSDI2018Yeo,INFOCOM2020Dasari}. 
Specifically, \textit{Dejavu} in \cite{HotMobile2019Hu} enhances the videoconferencing in real-time by employing the historical sessions. 
The deep neural network for the SR is applied to the adaptive streaming system in \cite{OSDI2018Yeo}, and the authors of \cite{INFOCOM2020Dasari} show the quality enhancement of the 360-Degree video streaming by using the SR technique. 
However, the above studies have not considered the adaptive SR which controls the quality enhancement level depending on the buffer and power state of the receiver, and not optimized the delivery decisions (e.g., transmission power, computing power) together. 

As the number of residual blocks (i.e., hidden layers) of the neural network increases, the deep learning-based SR achieves better performance at the expense of the speed of SR processing.
In other words, the tradeoff between the performance and computation time (delay) is observed. 
The authors of \cite{TMC2020Yi} offloads the computational SR tasks to the cloud to avoid the excessive latency; however, bitrate adaptation of video streaming and optimization of content delivery and network states are not considered. 
The anytime neural network (ANN) \cite{ArXiv2018Hu} fundamentally controls this tradeoff by allowing a quick and coarse prediction results and refining it if the computational budget is available. 
We applied the concept of the ANN to the GAN-based SR method to control the tradeoff between SR performance and computational tasks. 


\subsection{Learning-Based ABR Streaming}

Recently, the learning-based adaptive quality selection method for ABR streaming has been extensively researched in \cite{SIGCOMM2017Mao,GC2019Meng,RLRLW2019Mao}, but 
it does not include the characteristics of wireless networks.
After that, deep neural networks (DNNs) are used for jointly optimizing quality adaptation and resource allocation for dynamic streaming in wireless networks in \cite{Mobihoc2019Bhattacharyya,TCOMM2019Ye,INFOCOM2020Guan}.
In \cite{Mobihoc2019Bhattacharyya}, \textit{QFlow}, which is a reinforcement learning approach of selecting bitrates for wireless streaming by adaptively controlling flow assignments to queues is introduced.
Power-efficient wireless ABR streaming is proposed in \cite{TCOMM2019Ye} in which power control is jointly optimized with minimization of video transmission time by using deep reinforcement learning (DRL).
Furthermore, the DNN-assisted dynamic streaming using multi-path transmissions is presented in \cite{INFOCOM2020Guan}. 

In MEC-assisted streaming systems, the authors of \cite{TVT2020Guo} proposed the DRL method for quality adaptation and transcoding that balances the tradeoff between the QoE of video services and computational costs of transcoding. 
Similarly, a joint framework of quality adaptation and transcoding is presented in \cite{IoTJ2021Fu} using soft actor-critic DRL, which further reduces bitrate variance.

Nevertheless, quality enhancement at the edge device for smooth and high-quality streaming has not been widely studied yet, except for \cite{OSDI2018Yeo,HotNets2017Yeo,INFOCOM2020Zhang}. 
Quality enhancement of video contents using SR at the receiver side is first realized for adaptive video delivery in \cite{OSDI2018Yeo}, and further, the efficient content-aware video delivery is proposed by leveraging redundancy across videos in \cite{HotNets2017Yeo}; however, characteristics of wireless networks are not captured. 
In \cite{INFOCOM2020Zhang}, the DRL method of integrating the SR technique for quality improvement of videos with the wireless video streaming system is proposed. 
This scheme jointly pursues high quality, low quality variations, and infrequent rebuffer events; however, adaptive controls of computational tasks of transcoder and the DNN for SR are not considered.

\begin{table}[t!]
	\small
	\caption{System Description Parameters}
	\label{table:parameters}
	\begin{center}
		\scalebox{1}{
			\begin{tabular}{|l|l|}
				\hline
				$r(t)$ & Compression rate \\
				\hline
				$N(t)$ & Number of transmitting images \\
				\hline
				$P(t)$ & Transmit power \\
				\hline
				$d(t)$ & Depths of super-resolution network \\
				\hline 
				$u(t)$ & GPU core usage \\
				\hline
				$Q(t)$ & Transmitter queue length \\
				\hline
				$Z(t)$ & Receiver buffer length \\ 
				\hline
				$W(t)$ & Virtual queue for limiting power consumption \\
				\hline
				$\Theta(t)$ & Virtual queue for limiting GPU consumption \\
				\hline
				$\lambda(t)$ & Arrival rate of transmitter queue \\
				\hline
				$S(r(t))$ & Size of file compressed with $r(t)$ \\
				\hline
				$h(t)$ & Channel gain \\
				\hline
				$\tau(r, d, u)$ & Processing time of ASRGAN \\
				\hline
				$\Upsilon(r, d)$ & Task size of image recovery \\
				\hline
				$L$ & Number of possible quality levels \\
				\hline
				$t_0$ & discrete time duration \\
				\hline
				$\mathcal{B}$ & Bandwidth \\
				\hline
				$\eta$ & Threshold for average power consumption \\
				\hline
				$\xi$ & Threshold for average GPU core consumption \\
				\hline
				$P_0$ & Transmit power budget \\
				\hline
			\end{tabular}
		}
	\end{center}
\end{table}

\section{Problem Statement}
\label{sec:model}

This paper focuses on adaptive quality control in the wireless video streaming system as shown in Fig. \ref{fig:network_model}. 
Let the transmitter have the capability of compressing the high-quality images requested by a user and the user 
have the ability to enhance the quality of the received images.
Each video file or stream consist of a series of images, called as chunks that is in charge of the fixed playtime.
When the user starts to play the stream, the server delivers the desired video chunks in sequence to the user. 

\subsection{Transmitter Queue Model and Video Transcoding}

\begin{figure}[h]
	\centering
	\includegraphics[width=0.9\columnwidth]{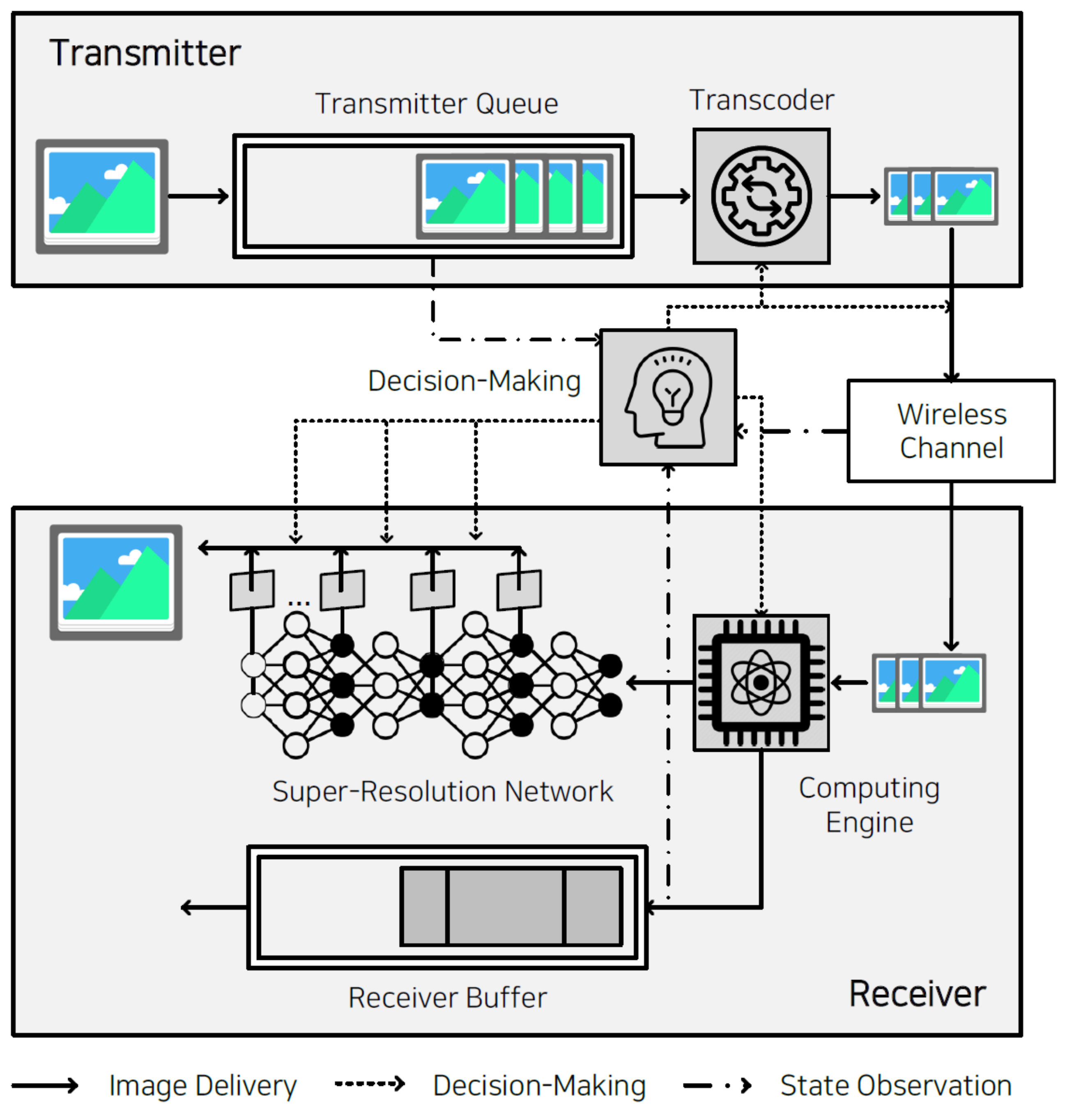}
	\caption{Adaptive Quality Control of Video Streaming System}
	\label{fig:network_model}
\end{figure}

Suppose that the transmitter has all of the desired video contents with the highest quality. 
These images are accumulated in the first-in-first-out (FIFO) transmitter queue as shown in Fig. \ref{fig:network_model}. 
The transmitter is deployed with the video transcoder; therefore, the desired images could be compressed before they are delivered to the user. 
Denote $\mathcal{R}=\{1,\cdots,N_r \}$ as the set of video bitrates, where $N_r$ is the number of possible video bitrates.
At the transmitter side, there are three decision parameters at every slot $t$ as follows: 1) the number of chunks supposed to deliver denoted by $N(t)$, 2) the transcoding rate of the images denoted by $r(t) \in \mathcal{R}$, and 3) the transmit power $P(t)$. 
Although the average streaming quality is very high, if the video quality is frequently fluctuating, it could degrade the user's QoS.
Therefore, assume that the chunks supposed to deliver at the same slot are compressed with the identical rate. 
In other words, $N(t)$ chunks have the identical transcoding rate $r(t)$.
The quality and the size of each chunk are determined by the transcoding rate $r(t)$. 
Denote $\mathcal{P}(r(t))$ and $S(r(t))$ as the quality and the size of the chunk compressed with the rate $r(t)$, respectively.

The queue dynamics in each time slot $t\in \{0,1,\cdots \}$ can be represented by $Q(t+1) = \max\{Q(t)-N(t) + \lambda(t), 0 \}$ and $Q(0) = 0$, where $Q(t)$ and $\lambda(t)$ stand for the queue backlog and the arrival process of the transmitter queue at slot $t$, respectively.
The queue backlog $Q(t)$ counts the number of chunks in the queue. $\lambda(t)$ and $N(t)$ semantically mean the numbers of the requested and transmitted images, respectively.
Simply, we assume the uniform distribution of $\lambda(t)$, i.e., $\lambda(t) \sim \mathcal{U}[0, \lambda_{\text{max}}]$, where $\lambda_{\text{max}}$ is the maximum number of the chunks that can be transmitted at each slot. 
On the other hand, $N(t)$ obviously depends on the capacity of the communication link between the server and the user and the transcoding rates of images as follows:
\begin{equation}
N(t) S(r(t)) \leq C(t) = t_0 \mathcal{B} \log_2 \Big( 1+ \frac{P(t) |h(t)|^2}{\sigma^2} \Big), \label{eq:rate}
\end{equation}
where $C(t)$ is the channel capacity in bits at slot $t$. 
Also, $t_0$ is the time duration of each slot, $\mathcal{B}$ is the bandwidth, $h(t)$ is the channel gain between the server and the user, and $\sigma^2$ is the noise variance.

The Rayleigh fading channel is assumed for the communication link from the server to the user.
Denote the channel with $h=\sqrt{L}g$, where $L=1/l^{\gamma}$ controls slow fading with $l$ being the server-user distance and $g$ represents the fast fading component having a complex Gaussian distribution, $g \sim CN(0,1)$.
Here, $\gamma$ is the pathloss exponent. 
Since the transmitter has the finite power budget $P_0$, i.e., $P(t) \leq P_0$, we can assume the upper bound on the expected value of $N(t)$, i.e., $\mathbb{E}[N(t)] \leq N_{\text{max}}$.

\subsection{Adaptive Super-Resolution Generative Adversarial Network (ASRGAN)}
\label{sec.DSRGAN} 

In this section, we introduce the Adaptive Super Resolution Generative Adversarial Network (ASRGAN). 
The ASRGAN is based on SRGAN~\cite{SRGAN2017} and is inspired from depth-controllable very deep super-resolution network (DCVDSR)~\cite{IJCNN2019Kim,ArXiv2018Hu}. 
Denote the lossless high resolution image as $I^{HR}$ and the compression image with the compressed rate $r$ as $I^{LR}(r)$. 
The ASRGAN consists of two components; the generator $G$ and the discriminator $D$. 
$K$ residual blocks in $G$ enhance the quality of $I^{LR}(r)$.
In addition, the ASRGAN can extract the feature of the input image from every $k\tau$-th redisual block for all $k\in\mathbb{N}^1[1,K/\tau]$. 
The extracted feature of the $k\tau$-th residual block is defined as follows:
\begin{equation}
    I^{SR}_{k\tau}(r) = G_{\theta_{k\tau}}(I^{LR}(r)),
\end{equation}
where $\theta_{k\tau}$ stands for a set of parameters of all blocks from the initial one to the $k\tau$-th residual ones $\forall k \in \mathbb{N}^1[1,K/\tau] $, respectively.

\subsubsection{Loss function}
\label{sec.DSRGAN-LF} 

In this subsection, we introduce the loss function of the ASRGAN. 
According to~\cite{SRGAN2017}, the mean squared error (MSE) $L^{SR}_{MSE}$, the Euclidian distance loss $L^{SR}_{VGG}$, and the adversarial loss $L^{SR}_{Gen}$ are used for introducing a complicated loss function.
Here, $L^{SR}_{VGG}$ compares the feature of the pretrained VGG19 to $G$, and $L^{SR}_{Gen}$ represents the distinguishness whether the unknown image is the output of $G$ or the $I^{HR}${\color{blue}.} 
The aforementioned loss function represents as follows:
\begin{eqnarray}   
    L^{SR}_{MSE}&\triangleq&\parallel I^{HR} - I^{SR}\parallel_2\\
    L^{SR}_{VGG}&\triangleq&\parallel (\phi_{\theta_\phi}(I^{HR}) - \phi_{\theta_\phi}(I^{SR})\parallel_2\\
    L^{SR}_{Gen}&\triangleq&-\text{log}D_{\theta_D}(I^{SR}),
\end{eqnarray}
where $\phi$ represents VGG19, and $\parallel \cdot \parallel_2$ stands for the L2 loss. 
Since the ASRGAN can extract SR images $I^{SR}_{k\tau}$ from $G_{\theta_{k\tau}}$, the loss function should be designed in the consideration of training $G_{\theta_{k\tau}}$ for all $k$. 
Therefore, we propose that the loss functions (e.g., $L^{SR}_{MSE}$, $L^{SR}_{VGG}$ and $L^{SR}_{Gen}$) are reorganized into the form of the weighted sum. 
The newly introduced loss function is presented as follows:
\begin{eqnarray}   
L^{\star}_{MSE} & \triangleq & \sum\limits_{k=1}^{K/\tau} \delta_{k\tau} \parallel I^{HR} - I^{SR}_{k\tau}(r) \parallel_2 \label{eq:obj1}\\
L^{\star}_{VGG}&\triangleq& \sum\limits_{k=1}^{K/\tau} \delta_{k\tau} \parallel (\phi_{\theta_{\phi}}(I^{HR}) - \phi_{\theta_{\phi}}(I^{SR}_{k\tau}(r))\parallel_2 \label{eq:obj2}\\
L^{\star}_{Gen}&\triangleq& -\sum\limits_{k=1}^{K/\tau} \delta_{k\tau} \text{log}D_{\theta_D}(I^{SR}_k(r))\label{eq:obj3}\\
\text{where} & ~ &\sum\limits_{k=1}^{K/\tau} \delta_{k\tau} = 1 \\
            & ~ & \delta_{k\tau} \geq \delta_{\tau(k-1)} > 0,~~\forall k \in \mathbb{Z}^1[2,K/\tau]\label{eq:delta},
\end{eqnarray}
where $\delta_{k\tau}$ is the weight factor for the loss of $G_{\theta_{k\tau}}$. 
According to \cite{ArXiv2018Hu}, it is possible to train the neural network successfully, if the weight $\delta_{k\tau}$ for the shallow residual block is small (i.e., $k\tau$ is small) and the weight increases as the residual block goes deeper (i.e., $k\tau$ is large), which is described in \eqref{eq:delta}.

Finally, the generator parameter $G_{\theta_{k\tau}}$ and the discriminator parameter $\theta_D$ can be trained by minimizing \eqref{eq:obj1}--\eqref{eq:obj3} as follows:
\begin{equation}   
\underset{\theta_G,\theta_D}{\min}~ \zeta^{SR}_{M}L^{\star}_{MSE}+\zeta^{SR}_{V}L^{\star}_{VGG}+\zeta^{SR}_{G}L^{\star}_{Gen},
\end{equation}
where $\zeta^{SR}_{M}$, $\zeta^{SR}_{V}$ and $\zeta^{SR}_{G}$ stand for the scaling factor for $L^{\star}_{MSE}$, $L^{\star}_{VGG}$ and $L^{\star}_{Gen}$, respectively. 

\subsubsection{Data preprocessing}
\label{sec.DSRGAN-DP}

In this subsection, we introduce the method of data preprocessing for training ASRGAN. 
The transmitter has high resolution image $I^{HR}$ of size $W\times H\times 3$. 
The transmitter compress $I^{HR}$ into $I^{LR}(r)$ of size $W/r \times H/r \times 3$ with bicubic interpolation. Subsequently, the transmitter send $I^{LR}(r)$ to the receiver. Meanwhile, the receiver transforms the compressed image $I^{LR}(r)$ into image of size $W \times H \times 3$. Regardless of the resolution, $I^{LR}(r)$ has $W\times H \times 3$ size for all $r$ and ASRGAN is able to train without concerning additional model against the image size and resolution.

\subsection{Receiver Buffer Model and Video Quality Enhancement}

The user device is deployed with a DNN that can enhance the quality of the compressed video chunks by using the SR technique. 
Here, we adopt the ASRGAN explained in Sec \ref{sec.DSRGAN-LF} for SR operation.
The ASRGAN can dynamically select the depth of the neural network to be used for SR with only minimal additional parameters. 
Obviously, the larger depth leads 
to a better quality of the resulting images; however, the longer processing time is required, and vice versa. 
Based on the ASRGAN, the two decisions should be made at the user sides: 1) the number of depths to be used for SR denoted by $d(t)$, and 2) the number of CPU cores to be used for operating the ASRGAN denoted by $u(t)$. 

Since $N(t)$ is carefully determined depending on the channel capacity, we assume that $N(t)$ images can be successfully delivered to the user within the time duration of $[t,t+1)$.
The user device decides the appropriate $d(t)$ and $u(t)$ to operate ASRGAN for $N(t)$ received images.
Suppose that the identical depths and CPU cores are employed for enhancing the quality of $N(t)$ images.

In order to measure the computing time required for SR, we consider the virtual receiver buffer in which the processing time for improving the quality of $N(t)$ images by using the ASRGAN with the depth $d(t)$ and $u(t)$ CPU cores. 
The buffer dynamics in each time slot $t\in \{0,1,\cdots \}$ can be represented by $Z(t+1) = \max\{Z(t) + a(t) - t_0, 0 \}$ and $Z(0) = 0$, where $a(t)$ is the arrival process of the receiver buffer $Z(t)$.
The departure rate is a constant $t_0$ obviously, because $t_0$ goes by between two consecutive slots.
The arrival process $a(t)$ can be described by
\begin{align}
a(t) &= N(t) \cdot \tau \big( r(t), d(t), u(t) \big) \\
&= N(t) \cdot \mathcal{I}\big\{ u(t) \neq 0 \big\} \frac{\Upsilon(r(t), d(t)) \omega}{u(t)}, 
\end{align}
where $\Upsilon(r(t), d(t))$ is the task size (e.g., required CPU clocks) for enhancing the quality of the received chunks transcoded with the rate $r(t)$ by operating the ASRGAN with $d(t)$, and $\tau(r(t), d(t), u(t))$ is the processing time of the ASRGAN using $u(t)$ CPU cores.
Also, $\mathcal{I}\{.\}$ is the indicator function, and $\omega$ is the average CPU clocks for processing a bit.
Since $\mathbb{E}[N(t)] \leq N_{\text{max}}$ and $\tau(r(t), d(t), u(t))$ is finite if $u(t) \neq 0$, and we can assume $\mathbb{E}[a(t)] \leq a_{\text{max}}$. 

After finishing the SR operation for the received $N(t)$ images, the final output quality is determined. 
Let $\mathcal{P}(r(t), d(t))$ be the quality measure of the chunks that is transcoded with the rate $r(t)$ at the transmitter and whose quality is enhanced by using the ASRGAN with the depth $d(t)$ at the receiver. 

\subsection{Tradeoff among queuing delay, buffering time, video quality, power consumption, and GPU usage}
\label{subsec:tradeoff}

The video delivery latency in our wireless streaming system consists of the queuing delay, the chunk processing delay (i.e., delay time required for SR operation), and the transmission time. 
Since we assume that $N(t)$ is appropriately determined to be successfully delivered within one slot under the channel capacity $C(t)$, the transmission time could be ignored compared to the queuing delay and the image processing delay.
In this model, the queuing delay is caused when video chunks are waiting for being departed from the transmitter queue.
If the channel condition is weak, the server cannot transmit large number of the high-quality chunks, and it could generate the queuing delays. 
In this case, it is better to compress chunks and to transmit many of them. 
In addition to this, the server should consider the power consumption.
If the battery power is not sufficient, then the number of transmitting images is limited and the queuing delay increases. 

Additionally, the chunk processing delay results from the operation of the SR technique, and the average chunk processing delay is proportional to the average length of the receiver buffer, similar to the transmitter queuing delay. 
If the received chunks are compressed with the high rate $r(t)$, the user should choose large $d(t)$ to pursue the high quality. 
Depending on $r(t)$ and $d(t)$, the task size is determined, and the task becomes heavy which causes the processing delay as the user enhances the quality of the image more. 
In this case, a choice of large $u(t)$ could limit the excessive processing delays; however, in order to support the multi-programming system at the edge device, usage of CPU cores has to be also limited. 
Accordingly, the chunk processing delay depends on $r(t)$, $d(t)$, and $u(t)$. 
Also, the streaming delay or playback stall that the user experiences is closely related to the average chunk processing delay. 
In general, the buffering time is given to the user for receiving initial parts of the stream before the video playback, and the user experiences the streaming delay if the average buffer length of $Z(t)$ is longer than the given buffering time. 
The buffering time analysis will be explained in detail in Sec. \ref{subsec:buffering}.

According to \cite{LittlesThm}, the average queuing delay is proportional to the average queue length; therefore, we can reduce both the transmitter queuing delay and the chunk processing delay (i.e., buffering time) by limiting the transmitter queue backlog and the receiver buffer length.
To this end, the Lyapunov optimization theory \cite{Lyapunov} proved that the time-average queue backlogs can be limited by pursuing strong stability of both transmitter and receiver queues as follows:
\begin{equation}
\underset{T \rightarrow \infty}{\lim} \frac{1}{T} \sum_{t=0}^T \mathbb{E} [Q(t)] < \infty ~\text{and}~ \underset{T \rightarrow \infty}{\lim} \frac{1}{T} \sum_{t=0}^T \mathbb{E} [Z(t)] < \infty. 
\label{eq:stability}
\end{equation}
Based on the Lyapunov optimization theory, the upper bound on the time-average queue length is also derived by using the algorithm which minimizes the Lyapunov drift \cite{Lyapunov} and finally the queuing delay and the chunk processing delay can be limited by achieving queue stability in \eqref{eq:stability}. 
In this respect, many delay-constrained transmission policies which limit the queueing delay by pursuing the queue stability have been proposed in \cite{TVT2018Choi,TWC2019Choi2}. 
In this paper, simulation results in Section \ref{sec:simulation} show that the queueing delay
can be reduced by ensuring \eqref{eq:stability}, i.e., strong stability of the
queueing system.

Note that the average quality of the received chunks depends on the decisions at the transmitter and receiver sides both. 
Therefore, decision parameters, i.e., $N(t)$, $r(t)$, $P(t)$, $d(t)$, and $u(t)$ are jointly optimized for pursuing the high video quality, limiting the queuing delay and the chunk processing delay, and saving the transmit power consumption and CPU usage. 
Among these performance metrics, we can observe a variety of tradeoffs.
First of all, the high-quality chunks require small $r$ and large $d$, which results in an increase of delays. 
The server can reduce the queuing delay while transmitting the high-quality images by consuming large transmit power. 
Similarly at the user side, the chunk processing time can be limited while pursuing high quality at the expense of large usage of CPU cores, i.e., $u(t)$.

Thus, decisions on $N(t)$, $r(t)$, $P(t)$, $d(t)$, and $u(t)$ have to be carefully made depending on the current channel condition, and both transmitter and receiver states. 
We can imagine that a central controller gathers the network information from both transmitter and receiver and dynamically controls the video delivery and its quality enhancement. 
In general, the streaming service provider is deployed with the powerful server; therefore, the scenario in which the transmitter (i.e., the server) can observe the channel state information, deliver the desired chunks with appropriate quality, and let the receiver know how many depths of the ASRGAN and CPU cores are required is possible.

\section{Joint Optimization of Dynamic Video Delivery and Quality Enhancement}
\label{sec:joint_opt}

This section introduces the joint optimization problem of dynamic image delivery and quality enhancement that pursues the high-quality video, the limited latency, and the efficient uses of transmit power and receiver CPU.
Also, the Lyapunov-based decision method for solving the problem is presented.

\subsection{Problem Formulation}
\label{subsec:prob_formula}

As explained in Section \ref{sec:model}, the proposed video delivery scheme jointly makes decisions on the number of transmitting chunks, the transcoding rate and the transmit power at the transmitter side, the number of depths of the ASRGAN, and the number of CPU cores at the receiver side in every time slot. 
  We suppose that the perfect channel state information (CSI) is known at the central controller or the transmitter.
After they observe their own queue and buffer states respectively, the decisions are made for pursuing the average image quality.
The joint optimization problem is described as follows:

\begin{align}   
\underset{\mathbf{N}, \mathbf{r}, \mathbf{P}, \mathbf{d}, \mathbf{u}}{\min}&~~ \underset{T\rightarrow \infty}{\lim} \frac{1}{T} \sum\limits_{t=1}^{T} \mathbb{E}\Big[ \Big( \bar{\mathcal{P}} - \mathcal{P}(r(t), d(t)) \Big) N(t) \Big] \label{eq:opt-obj}\\
\text{s.t.}&~~ \underset{T\rightarrow \infty}{\lim} \frac{1}{T} \sum\limits_{t=1}^{T} \mathbb{E} [Q(t)] < \infty, \label{eq:opt-queue} \\
&~~ \underset{T\rightarrow \infty}{\lim} \frac{1}{T} \sum\limits_{t=1}^{T} \mathbb{E} [Z(t)] < \infty, \label{eq:opt-buffer} \\
&~~ \underset{T\rightarrow \infty}{\lim} \frac{1}{T} \sum\limits_{t=1}^{T} \mathbb{E} [P(t)] \leq \eta, \label{eq:opt-avg_power} \\
&~~ \underset{T\rightarrow \infty}{\lim} \frac{1}{T} \sum\limits_{t=1}^{T} \mathbb{E} [u(t)] \leq \xi, \label{eq:opt-gpu} \\
&~~N(t) S(r(t)) \leq t_0 \mathcal{B} \log_2 \Big( 1+ \frac{P(t)|h(t)|^2}{\sigma^2} \Big) \label{eq:opt-rate} \\
&~~0 \leq P(t) \leq P_0 \label{eq:opt-power} \\
&~~r(t) \in \mathcal{R},~d(t) \in \mathcal{D},~u(t) \in \mathcal{U}, \label{eq:opt-set}
\end{align}
where $\bar{\mathcal{P}}$ is the maximum quality measure, $\eta$ is the threshold for the average transmit power, $\xi$ is the threshold for the average GPU usage, and $P_0$ is the power budget.
Also, $\mathcal{D}$ and $\mathcal{U}$ are the sets of available depths of the ASRGAN and the available CPU cores, respectively. 
Since we adaptively choose the number of transmitting and receiving chunks, the objective function in \eqref{eq:opt-obj} is the long-term time-averaged quality degradation of the received chunks.
Also, $\mathbf{N} = [N(0), N(1), \cdots, N(T)]$, and $\mathbf{r}$, $\mathbf{P}$, $\mathbf{d}$ and $\mathbf{u}$ are defined in a similar manner.
Specifically, the expectation of \eqref{eq:opt-obj}--\eqref{eq:opt-gpu} is with respect to random channel realizations. 
The constraints of \eqref{eq:opt-queue} and \eqref{eq:opt-buffer} are for limiting the queueing delay and the chunk processing delay. 
The transmit power consumption and usage of CPU cores are limited by the constraints of \eqref{eq:opt-avg_power} and \eqref{eq:opt-gpu}, respectively, and the constraint \eqref{eq:opt-rate} 
comes from \eqref{eq:rate}, which demonstrates that
decisions on $N(t)$ and $r(t)$ depend on the channel capacity. 

\subsection{Min-Drift-Plus-Penalty Algorithm}
\label{subsec:lyapunov}

The problem in \eqref{eq:opt-obj}--\eqref{eq:opt-set} can be solved by the theory of Lyapunov optimization \cite{Lyapunov}. 
We first transform the inequality constraints of \eqref{eq:opt-avg_power} and \eqref{eq:opt-gpu} into the forms of queue stability. 
Specifically, define the virtual queues $W(t)$ and $\Theta(t)$ with the following update equations:
\begin{align}
W(t+1) &= \max\{ W(t) -\eta + P(t), 0\} \label{eq:virtual_update1} \\
\Theta(t+1) &= \max\{ \Theta(t) -\xi + u(t), 0\}. \label{eq:virtual_update2}
\end{align}
The strong stability of the virtual queues $W(t)$ and $\Theta(t)$ push the average of $P(t)$ and $u(t)$ to be smaller than $\eta$ and $\xi$, respectively.

Let $\Xi(t) = [Q(t), Z(t), W(t), \Theta(t)]^T$ be a concatenated vector of the actual and virtual queue backlogs.
Define the quadratic Lyapunov function $L[\Xi(t)]$ as follows:
\begin{equation}
L[\Xi(t)] = \frac{1}{2} \Big\{ Q(t)^2 + k_z Z(t)^2 + k_w W(t)^2 + k_{\theta} \Theta(t)^2 \Big\},
\end{equation}
where $k_z$, $k_w$ and $k_{\theta}$ are scaling coefficients for $Z(t)$, $W(t)$, and $\Theta(t)$, respectively.
Then, let $\Delta(.)$ be a conditional quadratic Lyapunov drift on $t$ that is formulated as $\mathbb{E}[L(\Xi(t+1)) - L(\Xi(t)) | \Xi(t)]$. 
According to Lyapunov optimization theory \cite{Lyapunov}, if we suppose that the transmitter or the central controller could observe the current queue state $\Xi(t)$, the dynamic policy achieving stability of the queues in \eqref{eq:opt-queue}--\eqref{eq:opt-buffer} and \eqref{eq:virtual_update1}--\eqref{eq:virtual_update2} can be designed by minimizing an upper bound on \textit{drift-plus-penalty} which is given by
\begin{equation}
\Delta(\Xi(t)) + V \cdot \mathbb{E} \Big[ \Big( \bar{\mathcal{P}} - \mathcal{P}(r(t), d(t)) \Big) N(t) \Big], \label{eq:dpp}
\end{equation}
where $V$ is a system parameter that gives a weight to the average video quality. 

Here, the upper bound on the Lyapunov drift can be obtained as 
\begin{align}
&L(\Xi(t+1)) - L(\Xi(t)) \nonumber \\
&~~= \frac{1}{2} \Big\{ Q(t+1)^2 - Q(t)^2 + k_z \big( Z(t+1)^2 - Z(t)^2 \big) \nonumber \\
&~~~~ + k_w \big( W(t+1)^2 - W(t)^2 \big) + k_{\theta} \big( \Theta(t+1)^2 - \Theta(t)^2 \big) \Big\} \nonumber \\
&~~\leq \frac{1}{2} \Big\{ N(t)^2 + \lambda(t)^2 + k_z \big( a(t)^2 + t_0^2 \big) \} \nonumber \\
&~~~~ + Q(t)(\lambda(t) - N(t)) + k_z Z(t)(a(t) - t_0) \nonumber \\
&~~~~ + \frac{1}{2} \Big\{ k_w (P(t) - \eta)^2 + k_{\theta} (u(t) - \xi)^2 \Big\} \nonumber \\
&~~~~ + k_w W(t)(P(t) - \eta) + k_{\theta} \Theta(t) (u(t) - \xi) \nonumber \\
&~~ \leq B + Q(t) (c - N(t)) + k_z Z(t) (a(t) - t_0) \nonumber \\
&~~~~ + k_w W(t) (P(t) - \eta) + k_{\theta} \Theta(t) (g(u(t)) - \xi),
\end{align}
where a constant $B$ is chosen to satisfy the following inequality:
\begin{equation}
\frac{1}{2} \Big\{ \lambda_{\text{max}}^2 + N_{\text{max}}^2 + a_{\text{max}}^2 + t_0^2 + \eta^2 + P_0^2 + \xi^2 + u_{\text{max}}^2 \Big\} \leq B.
\end{equation}

Then, the upper bound on the conditional Lyapunov drift is given by 
\begin{align}
\Delta(\Xi(t)) &= \mathbb{E} \Big[ L(\Xi(t+1)) - L(\Xi(t)) \Big| \Xi(t) \Big] \nonumber \\
&\leq B + \mathbb{E} \Big[ Q(t)(\lambda(t) - N(t)) \Big| Q(t) \Big] \nonumber \\
&~~~~+ \mathbb{E} \Big[ k_z Z(t) (a(t) - t_0) \Big| Z(t) \Big] \nonumber \\
&~~~~+ \mathbb{E} \Big[ k_w W(t) (P(t) - \eta) \Big| W(t) \Big] \nonumber \\
&~~~~+ \mathbb{E} \Big[ k_{\theta} \Theta(t) (u(t) - \xi) \Big| \Theta(t) \Big].
\end{align}
According to \eqref{eq:dpp}, minimizing a bound on drift-plus-penalty is consistent with minimizing
\begin{align}
&\mathbb{E}\Big[ Q(t)(\lambda(t) - N(t)) \Big| Q(t) \Big] + \mathbb{E} \Big[ k_z Z(t) a(t) \Big| Z(t) \Big] \nonumber \\
&~~+ \mathbb{E} \Big[ k_w W(t)P(t) \Big| W(t) \Big] + \mathbb{E} \Big[ k_{\theta} \Theta(t) u(t) \Big| \Theta(t) \Big] \label{eq:bound_dpp}
\end{align}
We now use the concept of opportunistically minimizing the expectations; therefore, \eqref{eq:bound_dpp} is minimized by the algorithm which observes the current queue state $\Xi(t)$ and chooses $\Psi(t) = \{ N(t), r(t), P(t), d(t), u(t) \}$ to minimize 
\begin{align}
&\mathcal{D}(N(t),r(t),P(t),d(t),u(t)) \nonumber \\
&~~= -Q(t) N(t) + k_z Z(t) N(t) \frac{L(r(t), d(t)) \omega}{u(t)} + k_w W(t) P(t) \nonumber \\
&~~~~~~+ k_{\theta} \Theta(t) u(t) + V \cdot N(t) \Big( \bar{\mathcal{P}} - \mathcal{P}(r(t), d(t))\Big). \label{eq:dpp_obj}
\end{align}
Thus, we can reformulate the long-term problem of \eqref{eq:opt-obj}--\eqref{eq:opt-set} into the opportunistic min-drift-plus-penalty problem at every slot $t$ as follows:
\begin{align}
&\underset{\Psi=\{N,r,P,d,u \}}{\min}~ \mathcal{D}( N, r, P, d, u ) \label{eq:optdpp_obj} \\ 
&~~\text{s.t.}~~ N S(r) \leq t_0 \mathcal{B} \log_2 \Big( 1+ \frac{P|h|^2}{\sigma^2} \Big) \label{eq:optdpp_rate} \\
&~~~~~~~~0\leq P \leq P_0 \label{eq:optdpp_power} \\
&~~~~~~~~r\in \mathcal{R},~d \in \mathcal{D}(r),~u \in \mathcal{U}. \label{eq:optdpp_set}
\end{align}
In \eqref{eq:optdpp_obj}--\eqref{eq:optdpp_set}, the dependency on the time slot $t$ is omitted for simplicity because the decisions are made independently at every time slot.

From \eqref{eq:optdpp_obj}--\eqref{eq:optdpp_set}, we can anticipate how the algorithm works.
When the transmitter queue backlog is excessively long, many chunks are waiting to be delivered; therefore, the system can make the decisions on transmitting more chunks (i.e., large $N$) by compressing chunks with the large rate $r$ and/or consuming large power $P$. 
In this case, when the receiver buffer is excessively large, the heavy computational tasks are accumulated so that the receiver could not operate the SR at the expense of quality degradation or it uses large CPU cores $u$ for operating the SR.
On the other hand, when the receiver buffer is almost empty, it can deal with computational tasks for SR; therefore, the large depth $d$ of the ASRGAN is chosen to enhance the quality of the received chunks, and we can expect that the user can experience the high-quality streaming even with the small number of CPU cores. 
Meanwhile, if the transmitter queue backlog is short, it does not have to deliver the large number of chunks so that it can save its power and the transmitted chunks do not need to be transecoded with the high rate. 
It means that the quantity of the computational tasks to provide the high-quality streaming to the user is not excessively large.


System parameter $V$ in \eqref{eq:dpp_obj} is a weight factor for the term
representing the measure of video quality degradation. 
The relative value of $V$ to remaining terms is important to control queue backlogs and quality measures at every time slot. 
The appropriate initial value of $V$ needs to be obtained experimentally because it depends on the channel environments, relationship between video quality and file size, constraints on performances (i.e., $\eta$ and $\xi$) and system coefficients $k_z$, $k_w$, $k_{\theta}$. 
Also, $V \geq 0$ should be satisfied. 
If $V <0$, the user prefers low-quality videos even when the large number of chunks have already arrived at the user queue. 
Moreover, in the case of $V = 0$, the user only aims at accumulating queue backlogs without consideration of video quality. 
On the other hand, when $V \rightarrow \infty$, users do not consider the queue state, and thus they just request the highest-quality files. 
$V$ can be regarded as the parameter to control the trade-off between image quality and playback delay.

\section{Adaptive Control Algorithm for Video Delivery and Computational Tasks for Quality Adaptation}
\label{sec:schemes}

This sections proposes the adaptive algorithm, which controls the video delivery of the transmitter and computational SR tasks of the receiver by solving the problem of \eqref{eq:optdpp_obj}--\eqref{eq:optdpp_set}. 
In addition, we analyze the buffering time required for the stable, smooth, and high-quality streaming system, and introduce the comparison techniques to fairly compare the results of the proposed scheme.

\subsection{Adaptive Quality Control of Stable, Smooth, and High-Quality Streaming}
\label{subsec:mdpp}

Since the expected queue length is proportional to the average queuing delay according to Little's theorem, if the quality of transmitting chunks and transmit power are given, it is advantageous for the transmitter to deliver as many chunks in its queue as possible. 
Therefore, the inequality constraint on the data rate in \eqref{eq:optdpp_rate} can be converted into the equality constraint.
Also, the constraint \eqref{eq:optdpp_rate} gives the relationship among $N$, $r$, and $P$; therefore, if $N(t) = N$ and $r(t) = r$ are given, according to \eqref{eq:optdpp_rate}, the optimal power is obtained as 
\begin{equation}
P^*(t) = \frac{\sigma^2}{|h(t)|^2} \Big(2^{\frac{N S(r)}{t_0 \mathcal{B}}} - 1 \Big). \label{eq:opt_power}
\end{equation}

There are still many decision parameters to be made, i.e., $N(t)$, $r(t)$, $d(t)$, and $u(t)$; therefore, we formulate the subproblem with respect to $N(t)$ and $u(t)$ by considering $r(t)$ and $d(t)$ as constants. 
If $r(t) = r$ and $d(t)=d$ are given, we can reformulate the problem of \eqref{eq:optdpp_obj}--\eqref{eq:optdpp_set} as follows: 
\begin{align}
&\underset{N(t), u(t)}{\min}~ k_w W(t) \frac{1}{\Gamma |h(t)|^2} \Big( 2^{\frac{N(t)S(r)}{t_0 \mathcal{B}}}-1 \Big) + k_{\theta} \Theta(t) u(t) \nonumber \\
&~~~~+ N(t) \Big[ k_z Z(t) \frac{L(r,d)}{u(t)} - Q(t) + V\cdot (\bar{\mathcal{P}} - \mathcal{P}(r,d)) \Big]. \label{eq:subprob}
\end{align}
Here, we denote the optimal decision parameters of \eqref{eq:subprob} by $N'(t)$ and $u'(t)$ without any constraints.
Then, the objective function of \eqref{eq:subprob} is convex with respect to both $N(t)$ and $u(t)$, and the above problem in \eqref{eq:subprob} can be easily solved by using Karush–Kuhn–Tucker(KKT) conditions.
Thus, the following proposition provides the optimal solution of the problem in \eqref{eq:subprob}. 

\begin{proposition}
	\label{proposition}
	When $r(t) = r$ and $d(t) = d$ are given and $P(t)$ follows \eqref{eq:opt_power}, the optimal $N'(t)$ and $u'(t)$ have to satisfy the following equations:
	\begin{itemize}
		\item If $k_z Z(t) \frac{L(r,d)}{u(t)} - Q(t) + V\cdot (\bar{\mathcal{P}} - \mathcal{P}(r,d)) \geq 0$:
		\begin{equation}
		N'(t) = 0,~u'(t) = 0  \label{proposition_cond1}
		\end{equation}
		
		\item If $k_z Z(t) \frac{L(r,d)}{u(t)} - Q(t) + V\cdot (\bar{\mathcal{P}} - \mathcal{P}(r,d)) < 0$:
		\begin{align}
		&k_w \frac{W(t)}{\Gamma |h(t)|^2} \frac{S(r)}{t_0 \mathcal{B}} \ln 2 \cdot 2^{\frac{S(r)}{t_0 \mathcal{B}}N'(t)} \nonumber \\ 
		&~~~~+ k_z Z(t) L(r,d) \sqrt{\frac{k_z Z(t) L(r,d) N'(t)}{k_{\theta}\Theta(t)}} \nonumber \\ 
		&~~~~+ V \cdot (\bar{\mathcal{P}} - \mathcal{P}(r,d)) - Q(t) = 0 \label{proposition_cond2_N} \\
		&u'(t) = \sqrt{\frac{k_z Z(t) L(r,d) N'(t)}{k_{\theta} \Theta(t)}} \label{proposition_cond2_u}
		\end{align}
	\end{itemize}
\end{proposition}

Since $u(t) \in \mathcal{U}=\{0,1,\cdots, u_{\text{max}} \}$ and $N(t)$ is a nonnegative integer, in Proposition \ref{proposition}, we have to compare boundary conditions depending on the value of $u'(t)$. 
If $0\leq u'(t) \leq u_{\text{max}}$, four boundary conditions given by possible combinations of $N^*(t) \in \{ \lfloor N'(t) \rfloor, \lceil N'(t) \rceil \}$ and $u^*(t) \in \{ \lfloor u'(t) \rfloor, \lceil u'(t) \rceil \}$ have to be compared. 
Meanwhile, if $u'(t) \geq u_{\text{max}}$, we compare three boundary conditions as follows: 1) $N^*(t) = 0$ and $u^*(t) = 0$, 2) $N^*(t) = \lfloor N'(t) \rfloor$ and $u^*(t) = u_{\text{max}}$, and 3) $N^*(t) = \lceil N'(t) \rceil$ and $u^*(t) = u_{\text{max}}$.
Thus, if $r(t) = r$ and $d(t) = d$ are already given, then the optimal $N(t)$ and $u(t)$ are obtained by using Proposition \ref{proposition} and comparing the above four boundary conditions. 
Then, in order to find the optimal solution of the problem in \eqref{eq:optdpp_obj}--\eqref{eq:optdpp_set}, we can greedily test all joint combinations of decisions on $r(t)$ and $d(t)$ and finally can obtain $\Psi^*(t) = \{ N^*(t), r^*(t), P^*(t), d^*(t), u^*(t) \}$. 
The details are given in Algorithm \ref{algo}.

\begin{algorithm}[t!]
	\caption{Adaptive Quality Control for Stable, Smooth, and High-Quality Streaming \label{algo}}
	\begin{algorithmic}[1]
		\Require{\\
			\begin{itemize}
				\item $T$: Time slots
				\item $t_0$: Time slot duration
				\item $c$: Transmission rate
				\item $\eta$: Threshold for consumption of average transmit power
				\item $\xi$: Threshold for usage of CPU cores
				\item $\bar{Q}$: Maximum queue length
				\item $V$: Lyapunov coefficient
			\end{itemize}
		}
		\State{Initialization: $Q(0)=0$, $Z(0)=0$, $W(0)=0$, $V(0)=0$, and $\mathcal{D}^* = 10^{10}$.} 
		
		\For{$t \in \{0, 1,\cdots,T \}$}
		\For{$r^*(t) \in \mathcal{R}$ and $d^*(t) \in \mathcal{D}(r)$}
		\State{Find $N'(t)$ and $u'(t)$ according to Proposition \ref{proposition}.}
		\State{Compare four boundary conditions (i.e., $N^*(t) \in \{\lfloor N'(t) \rfloor, \lceil N'(t) \rceil \}$ and $u^*(t) \in \{ \lfloor u'(t) \rfloor, \lceil u'(t) \rceil \}$) and pick one of them minimizing \eqref{eq:subprob}.}
		\State{$P^*(t) \leftarrow \Big(2^{\frac{N^*(t)S(r^*(t))}{t_0 \mathcal{B}}}-1 \Big)\cdot \frac{\sigma^2}{|h(t)|^2}$}
		\State{Compute $\mathcal{D}(\Psi^*(t))$ by using \eqref{eq:dpp_obj}.}
		\If{$\mathcal{D}(\Psi^*(t)) < \mathcal{D}^{\star}$}
		\\$~~~~~~~~~~~~~~\mathcal{D}^{\star} \leftarrow \mathcal{D}(\Psi^*(t))$
		\\$~~~~~~~~~~~~~~N^{\star}(t) \leftarrow N^*(t),~r^{\star}(t) \leftarrow r^*(t)$
		\\$~~~~~~~~~~~~~~P^{\star}(t) \leftarrow P^*(t),~d^{\star}(t) \leftarrow d^*(t),~u^{\star}(t) \leftarrow u^*(t)$
		\EndIf
		\EndFor
		\State{\textbf{Tx}: Transcode $N^{\star}(t)$ video chunks with the rate $r^{\star}(t)$.}
		\State{\textbf{Tx}: Transmit $N^{\star}(t)$ transcoded chunks with transmit power $P^{\star}(t)$.}
		\State{\textbf{Rx}: Enhance the quality of the received images by using ASRGAN with depth $d^{\star}(t)$ and $u^{\star}(t)$ CPU cores.}
		\State{$a(t) \leftarrow N(t) \tau(d(t), u(t))$}
		\State{$Q(t+1) \leftarrow \max\{Q(t)+ c - N(t), 0 \}$}
		\State{$Z(t+1) \leftarrow \max\{Z(t)+ a(t) - \tau, 0 \}$}
		\State{$W(t+1) \leftarrow \max\{W(t)-\eta + P(t), 0 \}$}
		\State{$V(t+1) \leftarrow \max\{V(t)-\xi + g(u(t)), 0 \}$}
		\EndFor
	\end{algorithmic}
\end{algorithm}

\subsection{Buffering Time Analysis}
\label{subsec:buffering}

Owing to the data rate constraint of \eqref{eq:opt-rate}, the delivery of $N(t)$ chunks at slot $t$ can be successfully completed within the slot duration.
Since the user enjoys the chunks after the SR operation is completed, the buffer length $Z(t)$ becomes the delay time that the user experiences.
For a representative example scenario, an online streaming service provides the user the desired chunks in sequence, and the buffering time is required for smooth playback at the beginning of the stream. 
When the buffering time $b$ is given, it can be said that the user does not experience the playback delay if $Z(t) \leq b$.
If we set the strict delay constraint $b$ given $r(t) = r$ and $d(t) = d$, the following inequality has to be satisfied:
\begin{equation}
a(t) = N(t) \frac{\Upsilon(r,d) \omega}{u(t)} \leq b+t_0 - Z(t). \label{eq:buffering_const}
\end{equation}
For the strict delay constraint $b$, the inequality in \eqref{eq:buffering_const} could be considered as the constraint of the problem in \eqref{eq:subprob}, and the problem is still convex, so it can also be solved by satisfying the KKT conditions. 

Since $N(t)$ and $u(t)$ are integers, it is almost impossible to satisfy the equality condition of \eqref{eq:buffering_const}; therefore, the KKT conditions are the same as before, except for the addition of \eqref{eq:buffering_const} so that Proposition \ref{proposition} is still the solution if the inequality of \eqref{eq:buffering_const} is satisfied. 
First, \eqref{proposition_cond1} satisfies the inequality of \eqref{eq:buffering_const}. 
Second, if we consider both \eqref{proposition_cond2_u} and \eqref{eq:buffering_const} together, the maximum bound on $N(t)$ is obtained as 
\begin{equation}
N(t) \leq N_{\text{max}} = \frac{ \big( b+t_0-Z(t) \big)^2 }{\Upsilon(r,d)\omega} \cdot \frac{k_z Z(t)}{k_{\theta} \Theta(t)}. \label{eq:N_max}
\end{equation}
Accordingly, if $N^*(t)$ obtained from \eqref{proposition_cond2_N} is larger than $N_{\text{max}}$, the following boundary conditions have to be compared:
\begin{itemize}
	\item If $0\leq u^*(t) \leq u_{\text{max}}$: 1) $N^*(t) = 0$ and $u^*(t) = 0$, 2) $N^*(t) = N_{\text{max}}$ and $u^*(t) = \lfloor u'(t) \rfloor$, and 3) $N^*(t) = N_{\text{max}}$ and $u^*(t) = \lceil u'(t) \rceil$. 
	
	\item If $u^*(t) > u_{\text{max}}$: 1) $N^*(t) = 0$ and $u^*(t) = 0$, and 2) $N^*(t) = N_{\text{max}}$ and $u^*(t) = u_{\text{max}}$.
\end{itemize}

According to \eqref{eq:N_max}, the sufficiently long buffering time $b$ allows the user to receive many chunks every time. 
In this case, the user can receive many chunks in advance of the video playback during the buffering time $b$. 
Meanwhile, when $Z(t) \approx b$, the user worries about playback stall or streaming delays; therefore, the transmitter is better to deliver very small number of chunks and the user would not enhance the quality of the received chunks (i.e., $d(t) = 0$).

\subsection{Comparison Techniques}
\label{subsec:comparison_tech}

Our proposed dynamic video delivery algorithm (i.e., Algorithm \ref{algo}) has two important features: 1) coordination between the transmitter and the receiver, and 2) adaptive depth control of the ASRGAN. 
In order to verify the advantages of the proposed scheme, two comparison techniques having only one of the above features are introduced in this section. 

\subsubsection{Comp1: Separate Optimization at Transmitter and Receiver Sides}

This comparison scheme does not allow coordination between the transmitter and the receiver so that decision parameters are not jointly derived. 
Since the Lyapunov optimization process in Section \ref{subsec:lyapunov} is developed with the assumption that decisions are jointly made by allowing the cooperation between the transmitter and the receiver, the optimization problem should be differently formulated for being separately optimized at transmitter and receiver sides.
At the transmitter side, $N(t)$, $r(t)$, and $P(t)$ are determined from the following problem:
\begin{align}
\underset{N(t), r(t), P(t)}{\min}&~~ \underset{T\rightarrow \infty}{\lim} \frac{1}{T} \sum\limits_{t=1}^{T} \mathbb{E} \Big[ \Big( \bar{\mathcal{P}} - \mathcal{P}(r(t)) \Big) N(t) \Big] \nonumber \\
\text{s.t.}&~~ \eqref{eq:opt-queue}, \eqref{eq:opt-avg_power}, \eqref{eq:opt-rate}, \eqref{eq:opt-power}, \eqref{eq:opt-set}. \label{eq:comp1_tx_opt}
\end{align}
Similar to the Lyapunov optimization process in Section \ref{subsec:lyapunov}, the min-drift-plus-penalty algorithm at the transmitter side can be derived as 
\begin{align}
\underset{N(t), r(t), P(t)}{\min}&~~ -Q(t)N(t) + k_w W(t) P(t) \nonumber \\
&~~~~~+ V \cdot \big( \bar{\mathcal{P}} - \mathcal{P}(r(t)) \big) N(t) \nonumber \\
\text{s.t.}&~~ \eqref{eq:opt-rate}, \eqref{eq:opt-power}, \eqref{eq:opt-set}. \label{eq:comp1_tx_dpp}
\end{align}
Here, the transmitter greedily finds the optimal $N(t)$, $r(t)$, and $P(t)$ that minimizes the objective function of the problem in \eqref{eq:comp1_tx_dpp}.
On the other hand, the receiver side makes the optimal decisions on $d(t)$ and $u(t)$ by using the following problem: 
\begin{align}
\underset{d(t), u(t)}{\min}&~~ \underset{T\rightarrow \infty}{\lim} \frac{1}{T} \sum\limits_{t=1}^{T} \mathbb{E} \Big[ \Big( \bar{\mathcal{P}} - \mathcal{P}(r(t), d(t)) \Big) N(t) \Big] \nonumber \\
\text{s.t.}&~~ \eqref{eq:opt-buffer}, \eqref{eq:opt-gpu}, \eqref{eq:opt-set} \label{eq:comp1_rx_opt}
\end{align}
and it can be converted to the following min-drift-plus-penalty algorithm:
\begin{align}
\underset{d(t), u(t)}{\min}&~~ k_z Z(t) N(t) \frac{L(r(t), d(t)) \omega}{u(t)} + k_{\theta} \Theta(t) u(t) \nonumber \\ 
&~~~~~+ V \cdot \big( \bar{\mathcal{P}} - \mathcal{P}(r(t), d(t)) \big) N(t) \nonumber \\
\text{s.t.}&~~ \eqref{eq:opt-set}. \label{eq:comp1_rx_dpp}
\end{align}
Similarly, we also assume that the receiver greedily finds the optimal solution (i.e., $d(t)$ and $u(t)$) of the problem in \eqref{eq:comp1_rx_dpp}. 

\subsubsection{Comp2: Non-Adaptive Super-Resolution Network}

This comparison scheme does not employ the ASRGAN; therefore, the SR network is optimized for the fixed depth number and the receiver cannot control computational tasks of the SR network depending on the quality of the received chunks and the buffer state. 
We assume that the maximum depth of the SR network is always used; therefore, `Comp2' solves the identical problem of \eqref{eq:optdpp_obj}--\eqref{eq:optdpp_set} but $d(t) = d_{\text{max}}$.
Note that it is not the simplified version of `Comp1' because usage of CPU cores is still jointly determined with the transmitter decisions (i.e., $N(t$, $r(t)$, and $P(t)$) and the SR network of `Comp2' is different from that of `Comp2'. 
Since `Comp1' needs to adaptively control the number of the depths of the neural network, its SR network is optimized for multiple outputs from the ends of different depths; however, the neural network of `Comp2' is optimized for the fixed number of depths only.

\begin{figure}[t!]
    \centering
    \includegraphics[width=1\columnwidth]{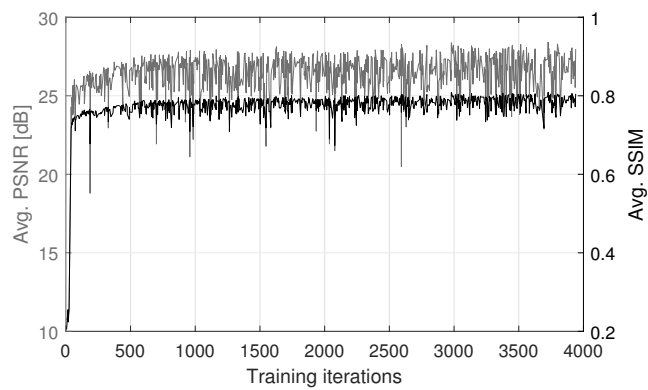}
    \caption{The transition of average PSNR and average SSIM by training.}
    \label{fig:psnrssim}
\end{figure}
\begin{figure*}
		\begin{tabular}{@{}p{0.0833\linewidth}@{}p{0.0833\linewidth}@{}p{0.0833\linewidth}@{}p{0.0833\linewidth}@{}p{0.0833\linewidth}@{}p{0.0833\linewidth}@{}|@{}p{0.0833\linewidth}@{}p{0.0833\linewidth}@{}p{0.0833\linewidth}@{}p{0.0833\linewidth}@{}p{0.0833\linewidth}@{}p{0.0833\linewidth}@{}}
		\centering Bicubic & \centering Depth 5 & \centering Depth 10 & \centering Depth 15 & \centering Depth 20 & \centering Depth 25 &
		\centering Bicubic & \centering Depth 5 & \centering Depth 10 & \centering Depth 15 & \centering Depth 20 & \centering Depth 25 \tabularnewline \toprule
		\includegraphics[page=1, width =1\linewidth]{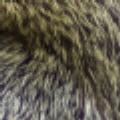}  &
		\includegraphics[page=1, width =1\linewidth]{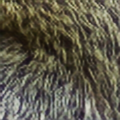} &
		\includegraphics[page=1, width =1\linewidth]{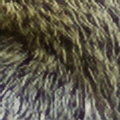} &
		\includegraphics[page=1, width =1\linewidth]{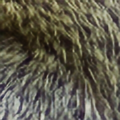} &
		\includegraphics[page=1, width =1\linewidth]{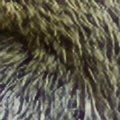} &
		\includegraphics[page=1, width =1\linewidth]{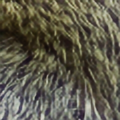} &
		\includegraphics[page=1, width =1\linewidth]{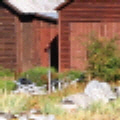}   &
		\includegraphics[page=1, width =1\linewidth]{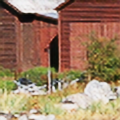} &
		\includegraphics[page=1, width =1\linewidth]{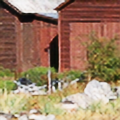} &
		\includegraphics[page=1, width =1\linewidth]{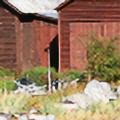} &
		\includegraphics[page=1, width =1\linewidth]{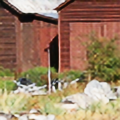} &
		\includegraphics[page=1, width =1\linewidth]{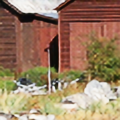} \tabularnewline
		\includegraphics[page=1, width =1\linewidth]{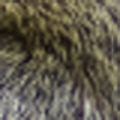}  &
		\includegraphics[page=1, width =1\linewidth]{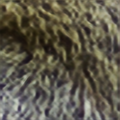} &
		\includegraphics[page=1, width =1\linewidth]{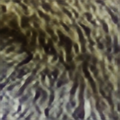} &
		\includegraphics[page=1, width =1\linewidth]{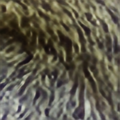} &
		\includegraphics[page=1, width =1\linewidth]{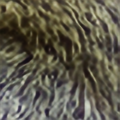} &
		\includegraphics[page=1, width =1\linewidth]{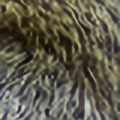} &
		\includegraphics[page=1, width =1\linewidth]{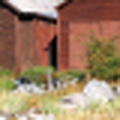}   &
		\includegraphics[page=1, width =1\linewidth]{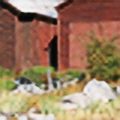} &
		\includegraphics[page=1, width =1\linewidth]{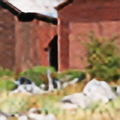} &
		\includegraphics[page=1, width =1\linewidth]{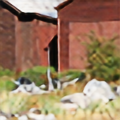} &
		\includegraphics[page=1, width =1\linewidth]{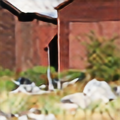} &
		\includegraphics[page=1, width =1\linewidth]{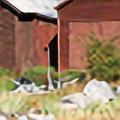} \tabularnewline
		
		\includegraphics[page=1, width =1\linewidth]{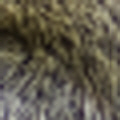}  &
		\includegraphics[page=1, width =1\linewidth]{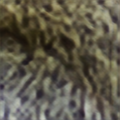} &
		\includegraphics[page=1, width =1\linewidth]{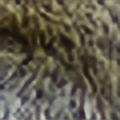} &
		\includegraphics[page=1, width =1\linewidth]{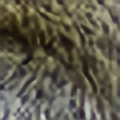} &
		\includegraphics[page=1, width =1\linewidth]{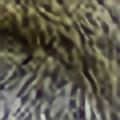} &
		\includegraphics[page=1, width =1\linewidth]{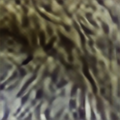} &
		\includegraphics[page=1, width =1\linewidth]{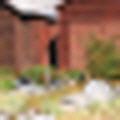}   &
		\includegraphics[page=1, width =1\linewidth]{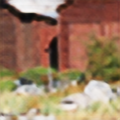} &
		\includegraphics[page=1, width =1\linewidth]{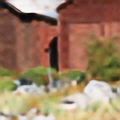} &
		\includegraphics[page=1, width =1\linewidth]{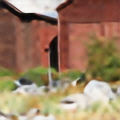} &
		\includegraphics[page=1, width =1\linewidth]{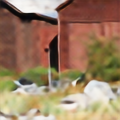} &
		\includegraphics[page=1, width =1\linewidth]{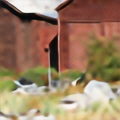}
		\tabularnewline
		\includegraphics[page=1, width =1\linewidth]{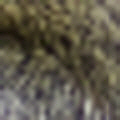}  &
		\includegraphics[page=1, width =1\linewidth]{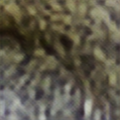} &
		\includegraphics[page=1, width =1\linewidth]{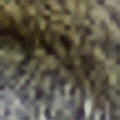} &
		\includegraphics[page=1, width =1\linewidth]{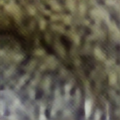} &
		\includegraphics[page=1, width =1\linewidth]{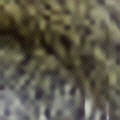} &
		\includegraphics[page=1, width =1\linewidth]{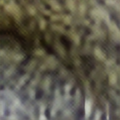} &
		\includegraphics[page=1, width =1\linewidth]{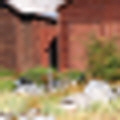}   &
		\includegraphics[page=1, width =1\linewidth]{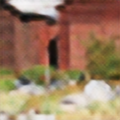} &
		\includegraphics[page=1, width =1\linewidth]{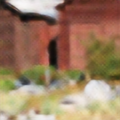} &
		\includegraphics[page=1, width =1\linewidth]{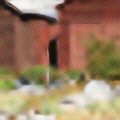} &
		\includegraphics[page=1, width =1\linewidth]{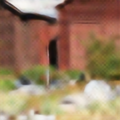} &
		\includegraphics[page=1, width =1\linewidth]{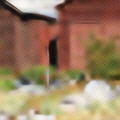} \tabularnewline
		\multicolumn{6}{c}{(a) Test image 1.} & \multicolumn{6}{c}{(b) Test image 2.}
        \tabularnewline			
		\includegraphics[page=1, width =1\linewidth]{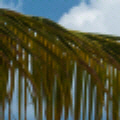}  &
		\includegraphics[page=1, width =1\linewidth]{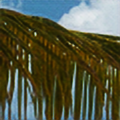} &
		\includegraphics[page=1, width =1\linewidth]{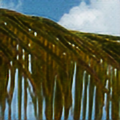} &
		\includegraphics[page=1, width =1\linewidth]{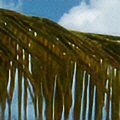} &
		\includegraphics[page=1, width =1\linewidth]{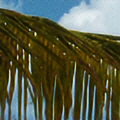} &
		\includegraphics[page=1, width =1\linewidth]{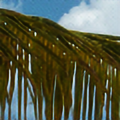} &
		\includegraphics[page=1, width =1\linewidth]{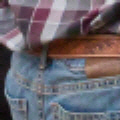}   &
		\includegraphics[page=1, width =1\linewidth]{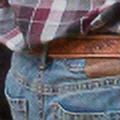} &
		\includegraphics[page=1, width =1\linewidth]{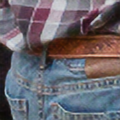} &
		\includegraphics[page=1, width =1\linewidth]{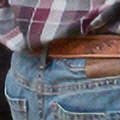} &
		\includegraphics[page=1, width =1\linewidth]{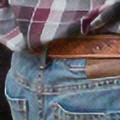} &
		\includegraphics[page=1, width =1\linewidth]{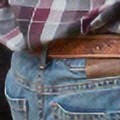}
		\tabularnewline
		\includegraphics[page=1, width =1\linewidth]{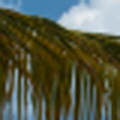}  &
		\includegraphics[page=1, width =1\linewidth]{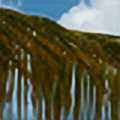} &
		\includegraphics[page=1, width =1\linewidth]{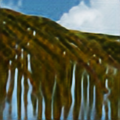} &
		\includegraphics[page=1, width =1\linewidth]{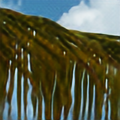} &
		\includegraphics[page=1, width =1\linewidth]{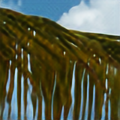} &
		\includegraphics[page=1, width =1\linewidth]{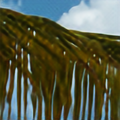} &
		\includegraphics[page=1, width =1\linewidth]{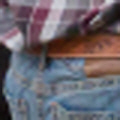}   &
		\includegraphics[page=1, width =1\linewidth]{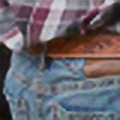} &
		\includegraphics[page=1, width =1\linewidth]{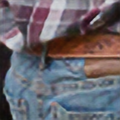} &
		\includegraphics[page=1, width =1\linewidth]{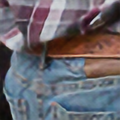} &
		\includegraphics[page=1, width =1\linewidth]{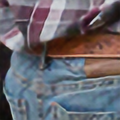} &
		\includegraphics[page=1, width =1\linewidth]{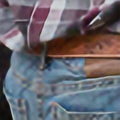} \tabularnewline
		\includegraphics[page=1, width =1\linewidth]{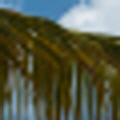}  &
		\includegraphics[page=1, width =1\linewidth]{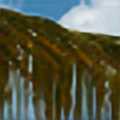} &
		\includegraphics[page=1, width =1\linewidth]{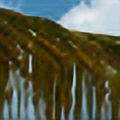} &
		\includegraphics[page=1, width =1\linewidth]{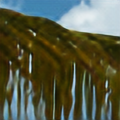} &
		\includegraphics[page=1, width =1\linewidth]{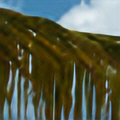} &
		\includegraphics[page=1, width =1\linewidth]{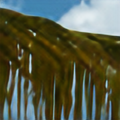} &
		\includegraphics[page=1, width =1\linewidth]{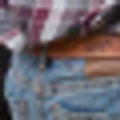}   &
		\includegraphics[page=1, width =1\linewidth]{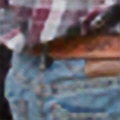} &
		\includegraphics[page=1, width =1\linewidth]{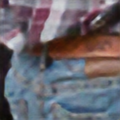} &
		\includegraphics[page=1, width =1\linewidth]{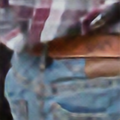} &
		\includegraphics[page=1, width =1\linewidth]{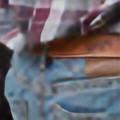} &
		\includegraphics[page=1, width =1\linewidth]{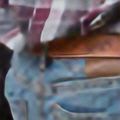} \tabularnewline
		\includegraphics[page=1, width =1\linewidth]{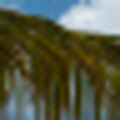}  &
		\includegraphics[page=1, width =1\linewidth]{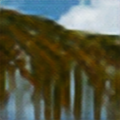} &
		\includegraphics[page=1, width =1\linewidth]{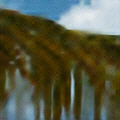} &
		\includegraphics[page=1, width =1\linewidth]{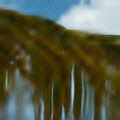} &
		\includegraphics[page=1, width =1\linewidth]{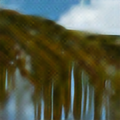} &
		\includegraphics[page=1, width =1\linewidth]{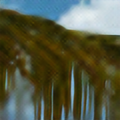} &
		\includegraphics[page=1, width =1\linewidth]{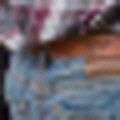}   &
		\includegraphics[page=1, width =1\linewidth]{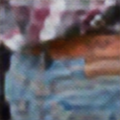} &
		\includegraphics[page=1, width =1\linewidth]{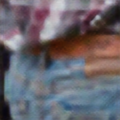} &
		\includegraphics[page=1, width =1\linewidth]{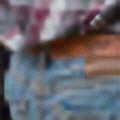} &
		\includegraphics[page=1, width =1\linewidth]{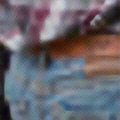} &
		\includegraphics[page=1, width =1\linewidth]{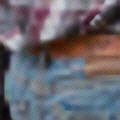} \tabularnewline
		\multicolumn{6}{c}{(c) Test image 3.} & \multicolumn{6}{c}{(d) Test image 4.}
        \tabularnewline	
		\end{tabular}
		\caption{It shows the inference image via the proposed ASRGAN for adaptive SR. For each of the figures (a/b/c/d), the first row is compressed with the ratio $r = 2$, the second row is compressed with the ratio $r = 3$, the third row is compressed with the $r = 4$, and the fourth row is compressed with the ratio $r = 5$.}
		\label{fig:testimages}
\end{figure*}

\section{Performance Evaluation}
\label{sec:simulation}

This section verifies the advantages of the proposed delay-sensitive and power-efficient quality control of dynamic video streaming compared to comparison techniques introduced in Section \ref{subsec:comparison_tech}. 
We first show the reliability of the adaptive SR whose performance is evaluated by observing PSNRs of output images. 
Next, the proposed scheme is shown to provide the sufficient QoS for the streaming user by balancing the tradeoff among the average video quality, queuing delay, chunk processing delay, transmitter power consumption, and receiver CPU usage. 

\begin{table*}[t!]%
	
	\begin{center}
	\begin{tabular}{c||c|c|c|c|c|c}
		\toprule[1.3pt]
		 & Depth 0& Depth 5& Depth 10& Depth 15& Depth 20& Depth 25\tabularnewline
		\midrule[1.0pt]
		 \multicolumn{1}{l||}{(a) Compression rate $(r)$}  & \multicolumn{6}{c}{PSNR (dB)/SSIM} \tabularnewline
		\midrule 
		$r=2$ & 30.60/0.859 & 30.66/0.866 & 31.01/0.885 & 31.87/0.899 & 32.20/0.904 & 32.26/0.906 \tabularnewline
		$r=3$ & 27.45/0.749 & 27.51/0.773 & 28.42/0.795 & 29.08/0.816 & 29.39/0.823 & 29.45/0.826 \tabularnewline
		$r=4$ & 25.82/0.672 & 26.29/0.701 & 26.85/0.724 & 27.49/0.749 & 27.73/0.759 & 27.79/0.762 \tabularnewline
		$r=5$ & 24.81/0.620 & 25.22/0.641 & 25.74/0.663 & 26.23/0.688 & 26.46/0.698 & 26.54/0.702 \tabularnewline
		\midrule[1.0pt]
		 \multicolumn{1}{l||}{(b) Weight of depth ($\delta_{k\tau}$)} & - & 0.0165 & 0.0660 & 0.264 & 0.323 & 0.330 \tabularnewline \midrule[1.0pt]
		 \multicolumn{1}{l||}{(c) Clocks $[10^9]$}& - & 1.007 & 1.441 & 1.864 & 2.283 & 2.669 \tabularnewline \midrule[1.0pt]
		 \multicolumn{1}{l||}{(d) Inference time $[$ms$]$}& - & 8.6 & 12.3 & 15.9 & 19.4 & 22.8 \tabularnewline \midrule[1.0pt]
%
	\end{tabular}
	\end{center}
	\caption{Information of the trained ASRGAN}
	\label{tab:pe}
\end{table*}

\subsection{Adaptive Super-Resolution Network}
\label{subsec:simul_adaptive_SR}

Our proposed ASRGAN is trained with 4,000 iterations and 32 batches. The generator $G$ is the ASRGAN and the VGG19 is adopted as discriminator $D$. 
The DIV2K high resolution dataset is used to train and test the ASRGAN~\cite{DIV2K}. 
High-resolution images are preprocessed as explained in Sec.~\ref{sec.DSRGAN-LF}. 
The preprocessed dataset with various resolutions (e.g. $r=2$, $r=3$, $r=4$, and $r=5$) is randomly cropped with the size of $120 \times\ 120 \times 3$. 
The training set and batch contain all pre-processed images regardless of $r$. 
The training is conducted with parameters of $\zeta^{SR}_{M} = 3.03$, $\zeta^{SR}_{V} = 3.03$, $\zeta^{SR}_{G} = 10^{-2}$ and $\delta_k$ as shown in Tab.~\ref{tab:pe}(b).

As a training result, Fig.~\ref{fig:psnrssim} shows the average PSNR and average SSIM in the training phase. 
The average PSNR and average SSIM increase during training phase, and approach 27.97 dB and 0.8051 separately at the end of the training phase (4,000 iterations). 
Fig.~\ref{fig:testimages} shows output images of the ASRGAN depending on the depth of the ASRGAN and the compression rate of input images. 
The test images shown in Fig.~\ref{fig:testimages} are from DIV2K test dataset. 
We can see that when the feature is extracted from the deeper depth of the ASRGAN, output images have a better resolution. 

The information of the trained ASRGAN is described in Table \ref{tab:pe}. 
The quality measures (i.e., PSNR and SSIM) of output images of the ASRGAN depending on the depth and the compression rate, and the required weights (i.e., $\delta_{k\tau}$) and CPU clocks for operating the ASRGAN with different depths are given. 
We can see that as the ASRGAN extracts the feature from the deeper depth, the quality measures as well as the required CPU clocks increase regardless of the compression rate $r$. 
In conclusion, we can confirm that there is a tradeoff between the image quality and the computational task, and it can be controlled by adjusting the depth of the ASRGAN.

\label{subsec:sr_result}

%

\begin{figure*}
\begin{tabular}{p{0.3\linewidth}p{0.3\linewidth}p{0.3\linewidth}}
	\includegraphics[page=1, width =1\linewidth]{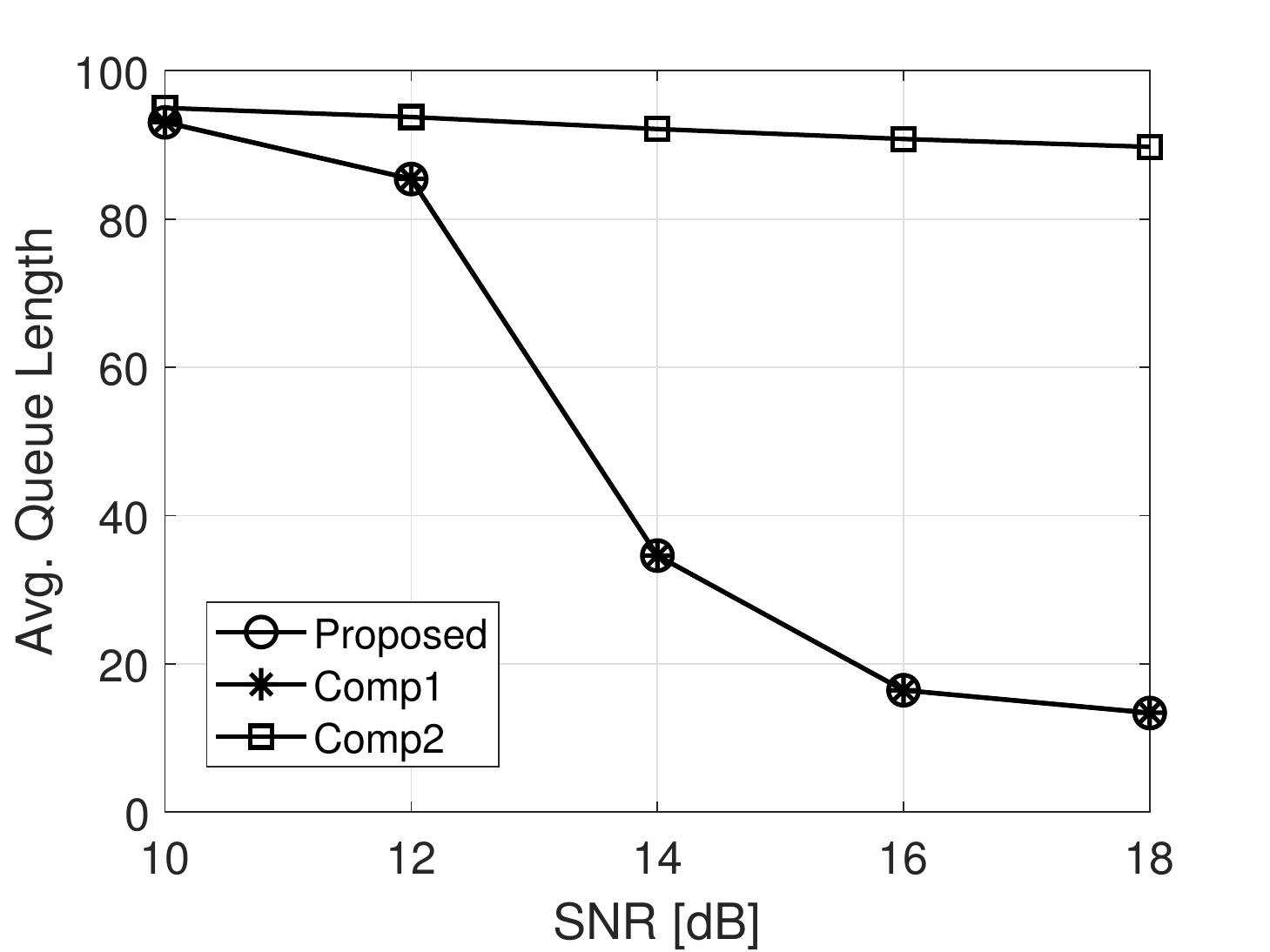} &
	\includegraphics[page=1, width =1\linewidth]{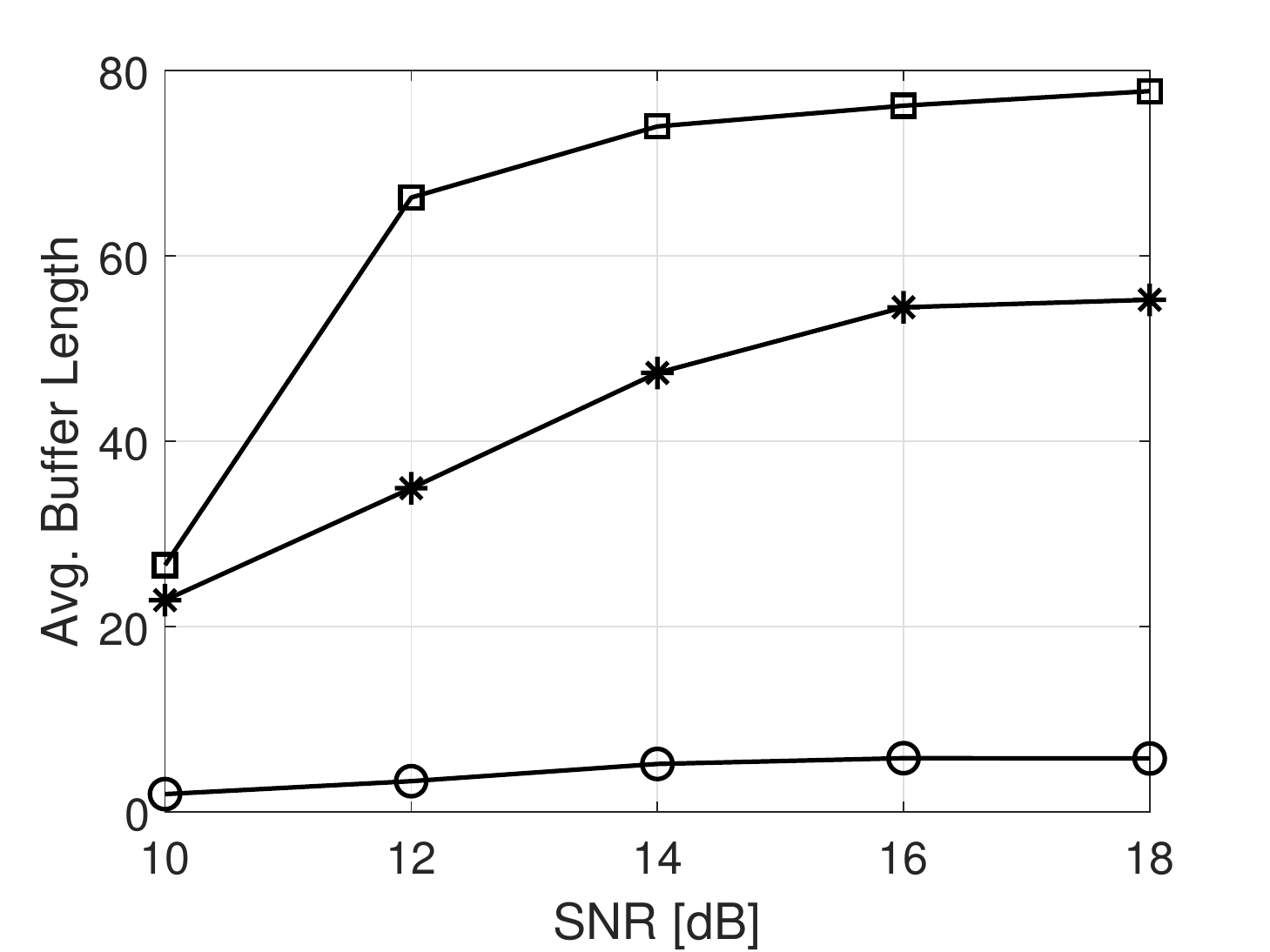} &
	\includegraphics[page=1, width =1\linewidth]{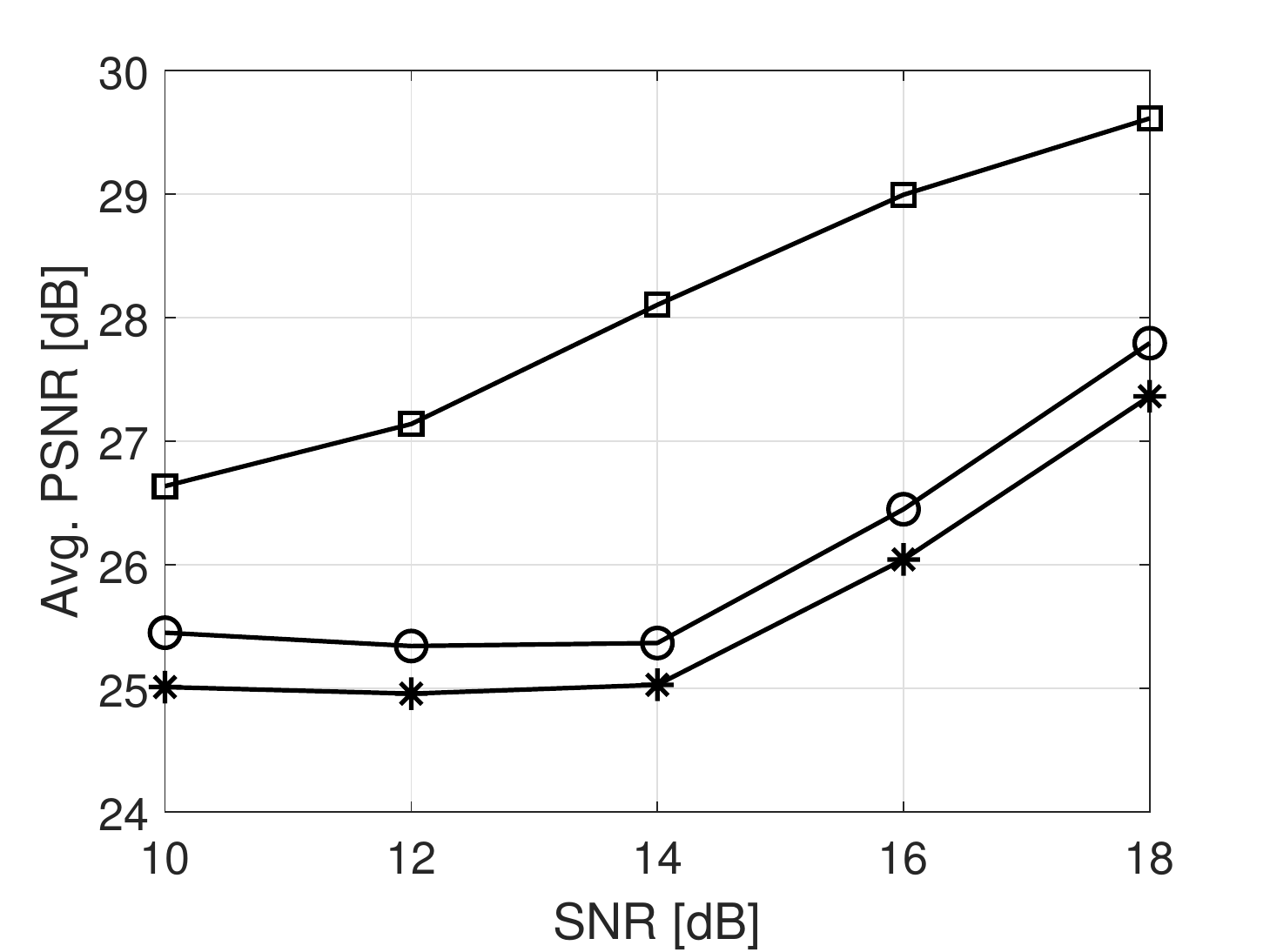}
     \tabularnewline
    \centering (a) Average Tx queue length &
    \centering (b) Average Rx buffer length &
    \centering (c) Average PSNR of received chunks
     \tabularnewline
	\includegraphics[page=1, width =1\linewidth]{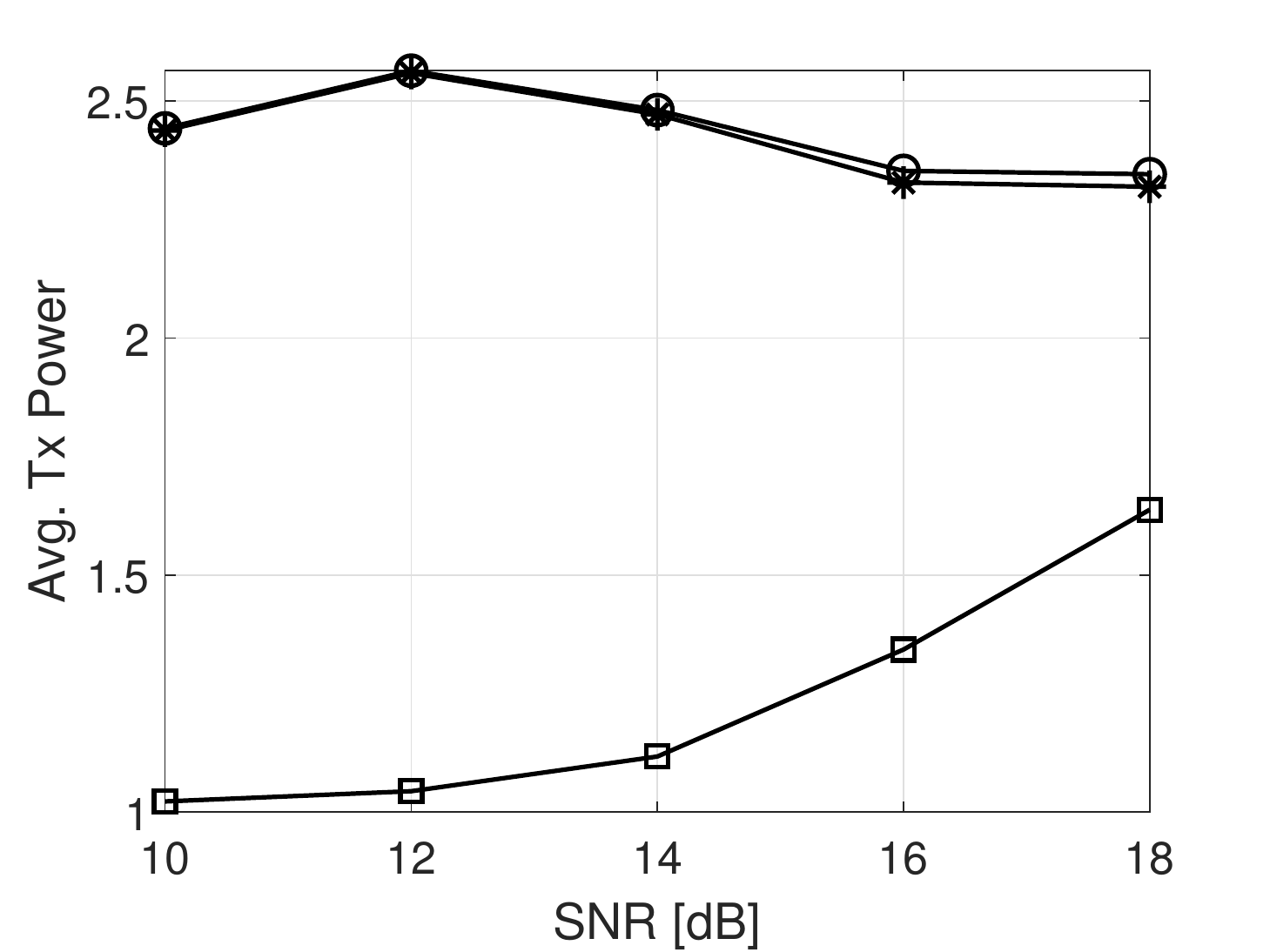} &
	\includegraphics[page=1, width =1\linewidth]{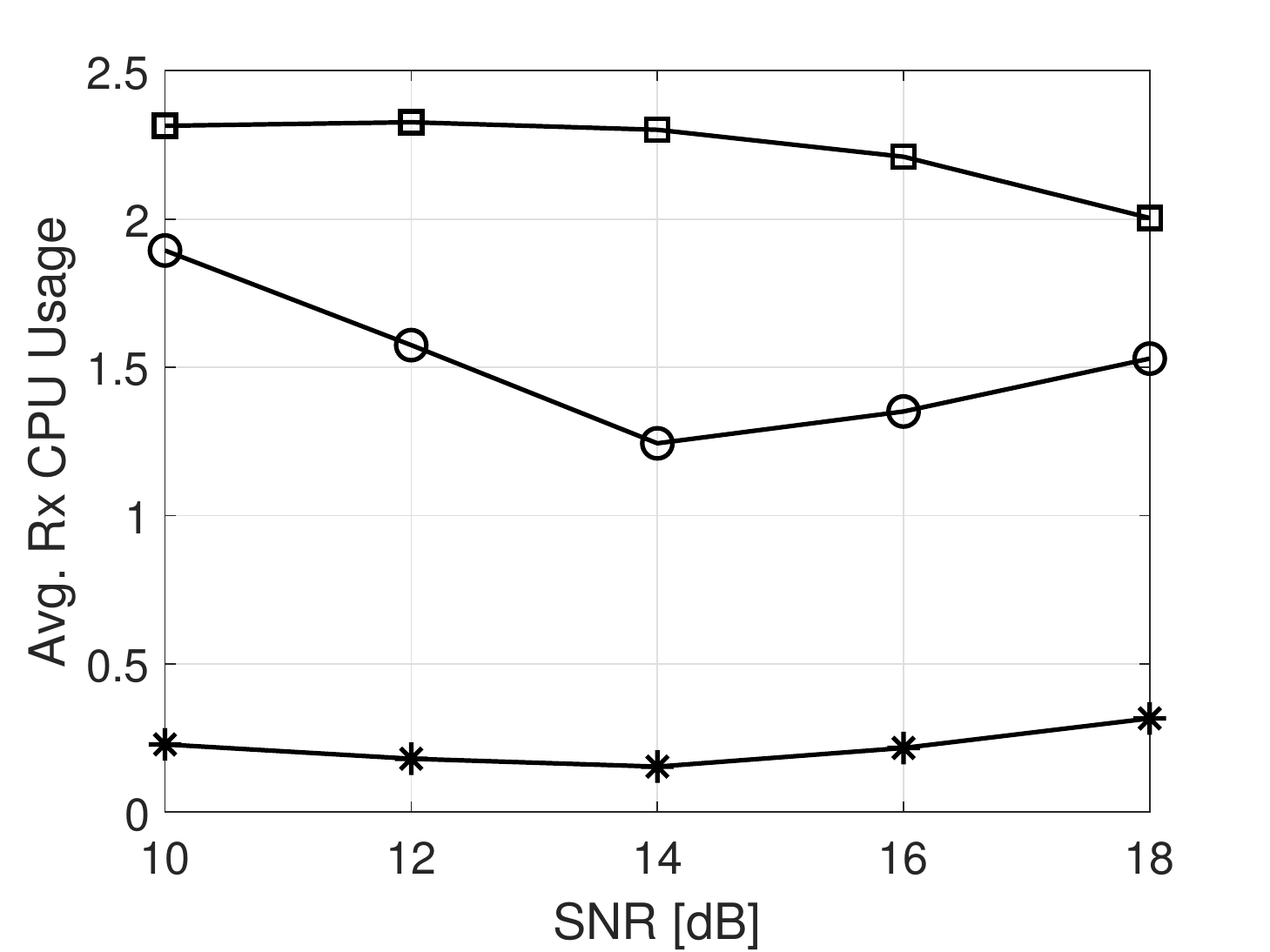} &
	\includegraphics[page=1, width =1\linewidth]{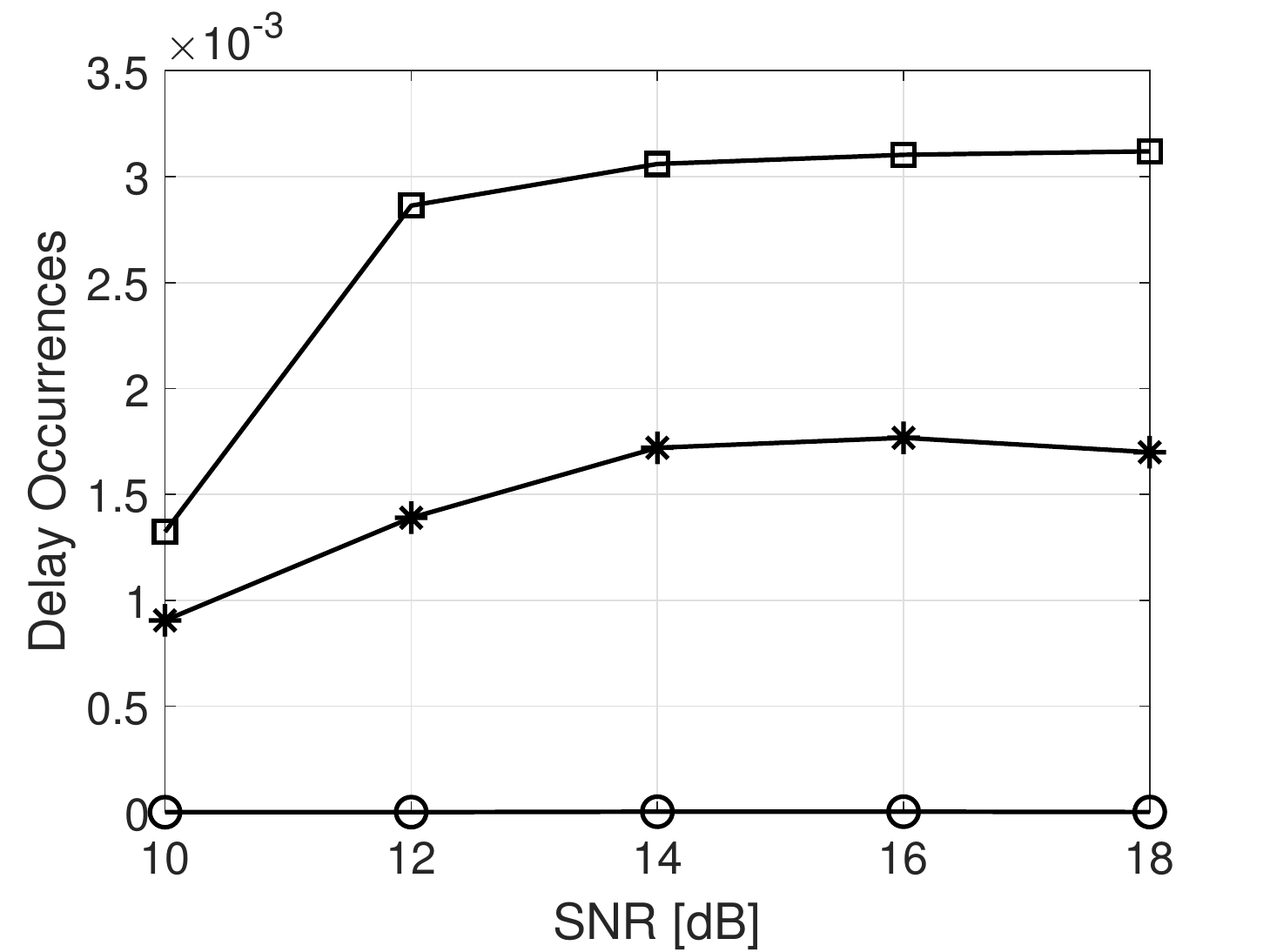}
     \tabularnewline
    \centering (d) Average Tx power consumption &
    \centering (e) Average Rx CPU usage &
    \centering (f) Delay occurrences
\end{tabular}
\caption{Performance metrics of the adaptive quality control of video streaming system vs. SNR}
\label{fig:result_snr}
\end{figure*}

\subsection{Numerical Results of Dynamic Video Streaming}
\label{subsec:numerical}

In this subsection, we experiment the real video streaming system using the ASRGAN whose specifications are explained in Section \ref{subsec:simul_adaptive_SR} and shown in Table \ref{tab:pe}. 
The system model for simulation is based on Fig. \ref{fig:network_model}, and SNR=16 dB and $V=0.01$ are assumed unless otherwise noted.
The constraints on the average power consumption and the average usage of CPU cores are $\eta = 2.5$ and $\xi = 2.5$, respectively.
In addition, $\mathcal{B} = 3$ MHz, $t_0 = 1$ sec, $N_{\textbf{max}} = 30$, $u_{\textbf{max}} = 10$, $\omega = 1.171$ GHz, $k_z = 1.0$, $k_w = 1.0$, and $k_{\theta} = 1.0$ are used.

Fig. \ref{fig:result_snr} shows the plots of various performance metrics of the streaming system versus the SNR. 
Overall, the proposed scheme balances all the performance metrics compared to other comparison techniques. 
Basically, if 'Comp2' determines to compress the chunks at the transmitter side, then it always operates SR using the SRGAN with the maximum depth without adaptive depth controls; therefore, its average video quality is better than other schemes using adaptive SR.
On the other hand, the average buffer length of `Comp2' is significantly longer than other schemes so that its delay incidence is excessively large and it diminishes the QoS of streaming users.

Since the transmitter and the receiver of `Comp1' separately control the transcoding rate and the SR operation in sequence, the receiver decisions (i.e., depths of the ASRGAN and CPU usage) are strictly dependent on the transmitter decisions. 
Meanwhile, the proposed scheme can jointly control the transmitter and the receiver decisions; therefore, we can obtain the better receiver decisions compared to `Comp1'. 
Accordingly, Fig. \ref{fig:result_snr} shows that the proposed scheme and `Comp1' provide almost the same results for the performance metrics related to the transmitter decisions (i.e., the average queue length and the transmitter power consumption); however, the proposed scheme greatly outperforms `Comp1' in terms of the performance metrics related to the receiver decisions (i.e., the average buffer length and the delay incidence). 
Especially, the long buffer length of `Comp1' leads the frequent playback stall events. 
Instead, the proposed scheme uses more CPU cores than `Comp1'; however, since $\xi=2.5$, it means that `Comp1' does not efficiently use the available CPU cores compared to the proposed scheme.
In addition, the proposed scheme provides better video quality to the user than `Comp1' because it jointly optimizes the decisions at the transmitter and the receiver sides. 
Thus, we can see that the proposed scheme balances the video quality and playback stall performances while achieving the limited power consumption and CPU usage.

\begin{figure*}
\begin{tabular}{p{0.3\linewidth}p{0.3\linewidth}p{0.3\linewidth}}
	\includegraphics[page=1, width =1\linewidth]{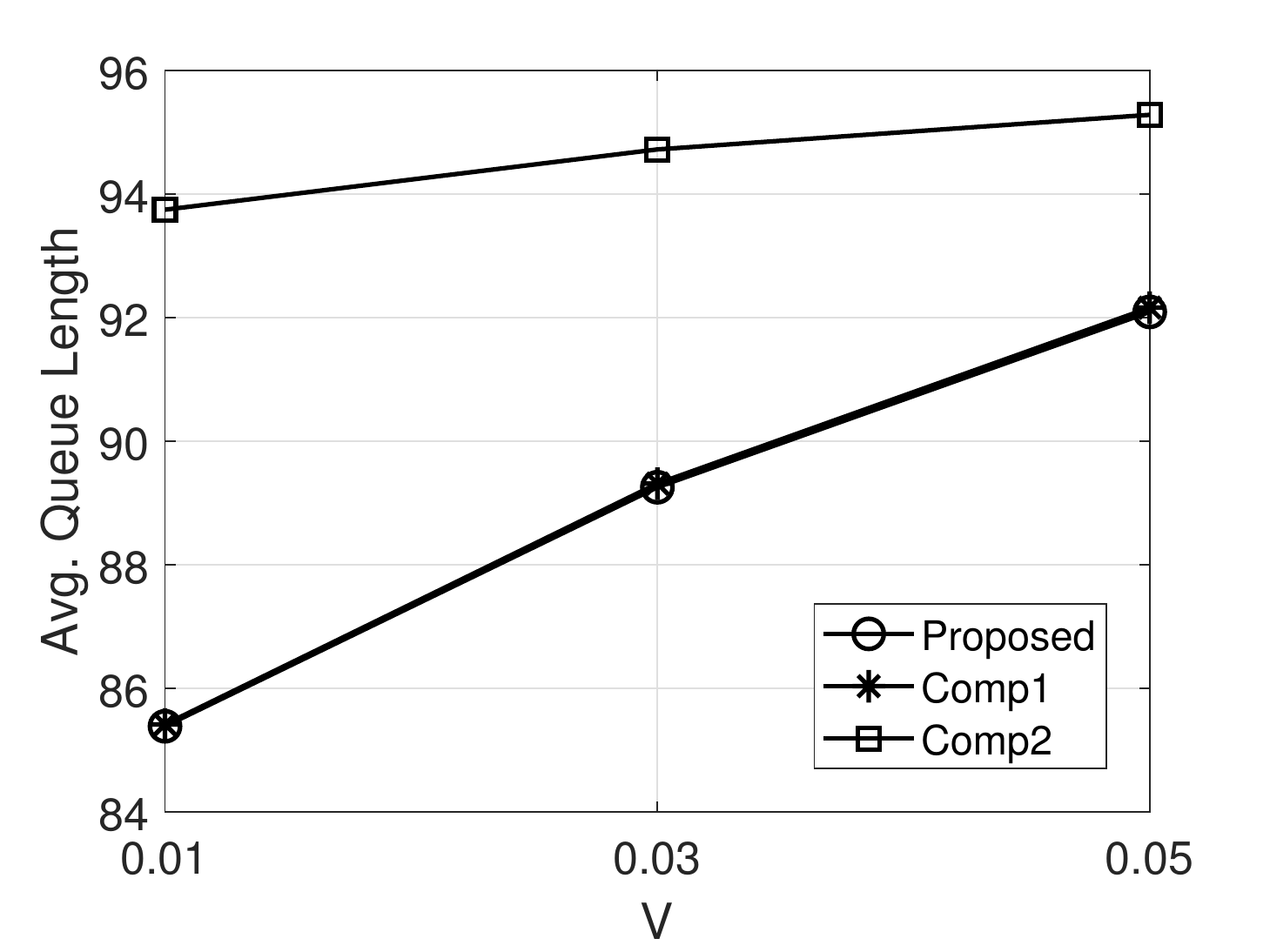} &
	\includegraphics[page=1, width =1\linewidth]{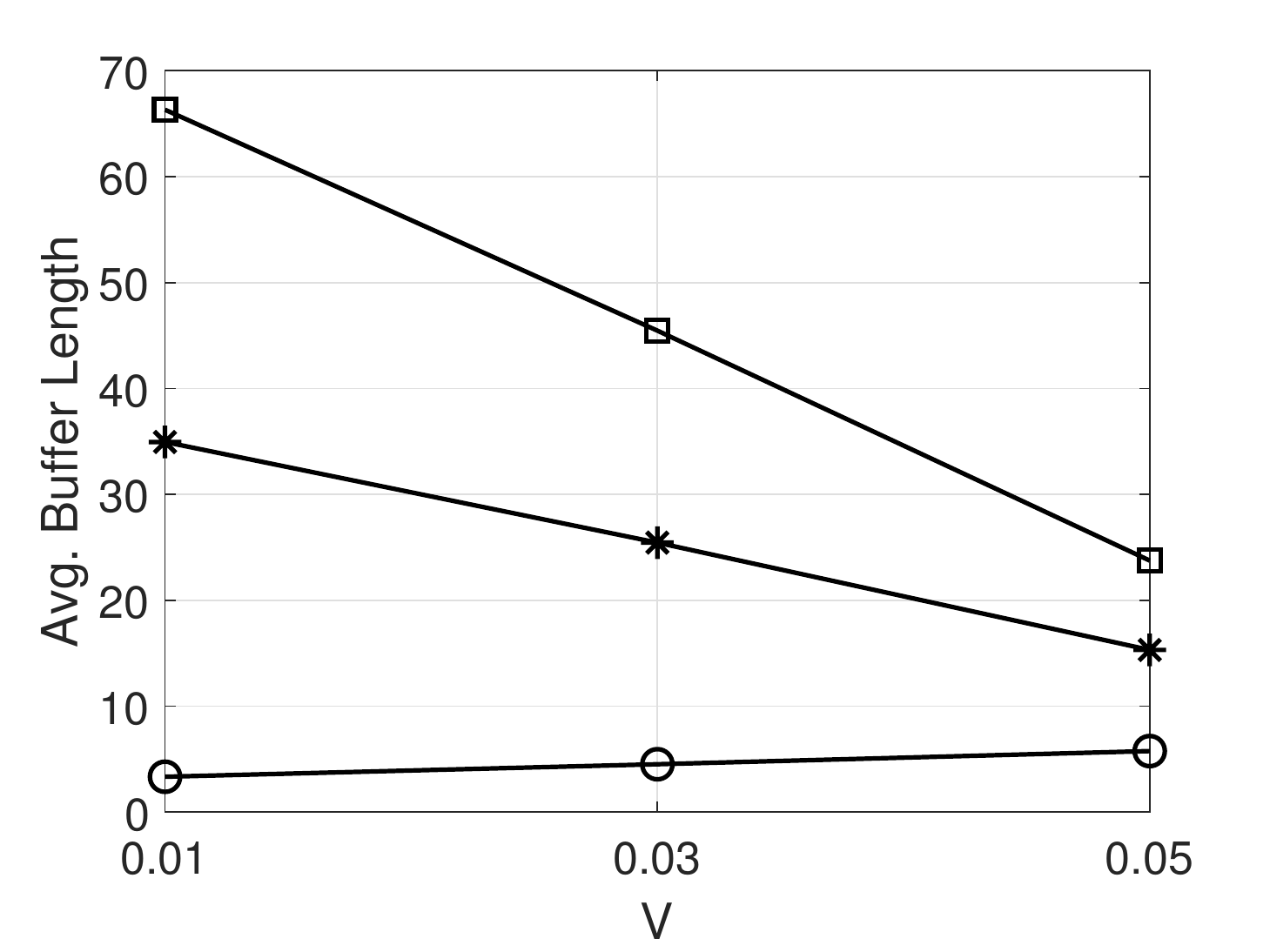} &
	\includegraphics[page=1, width =1\linewidth]{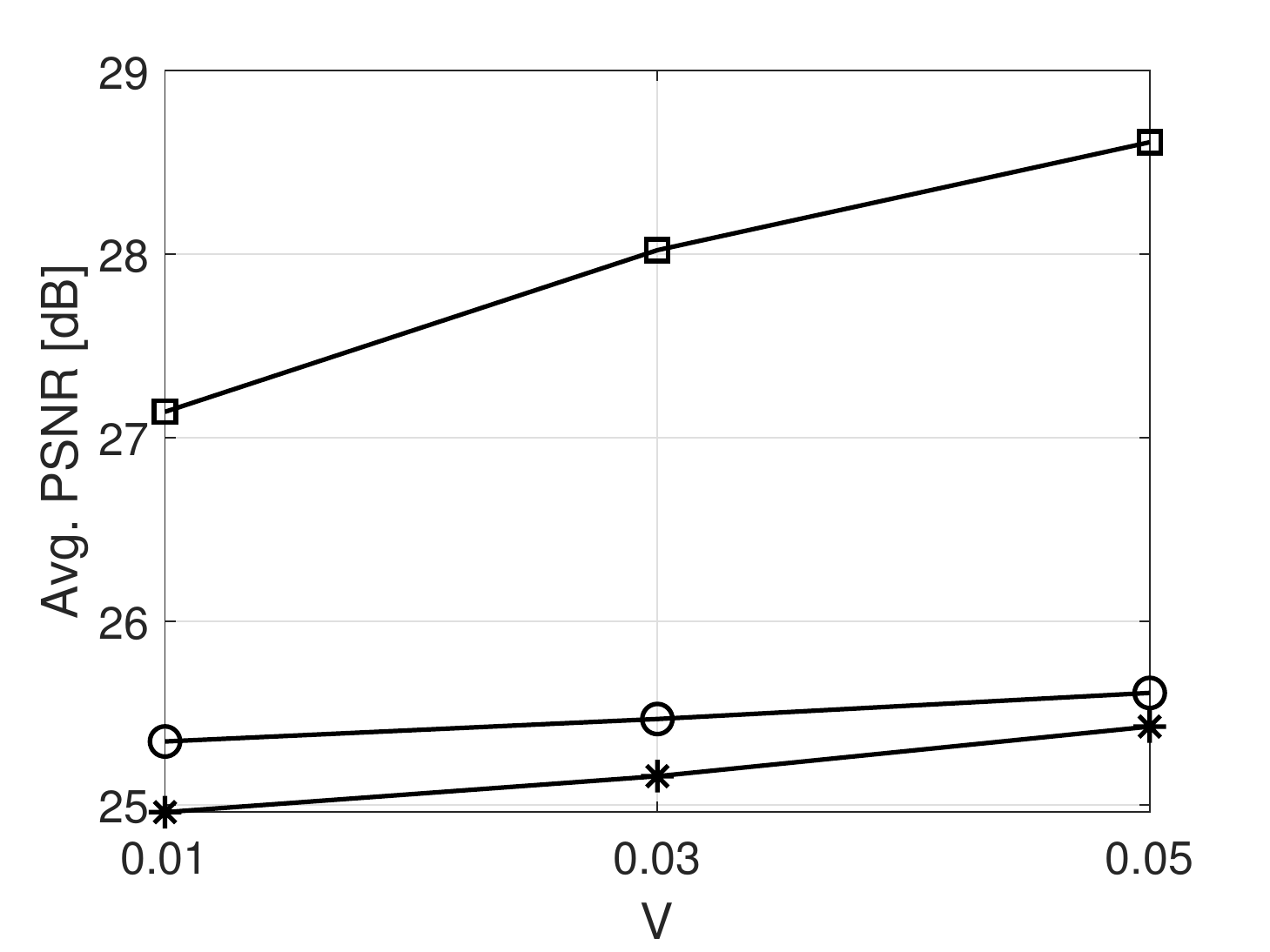}
     \tabularnewline
    \centering (a) Average Tx queue length &
    \centering (b) Average Rx buffer length &
    \centering (c) Average PSNR of received chunks
     \tabularnewline
	\includegraphics[page=1, width =1\linewidth]{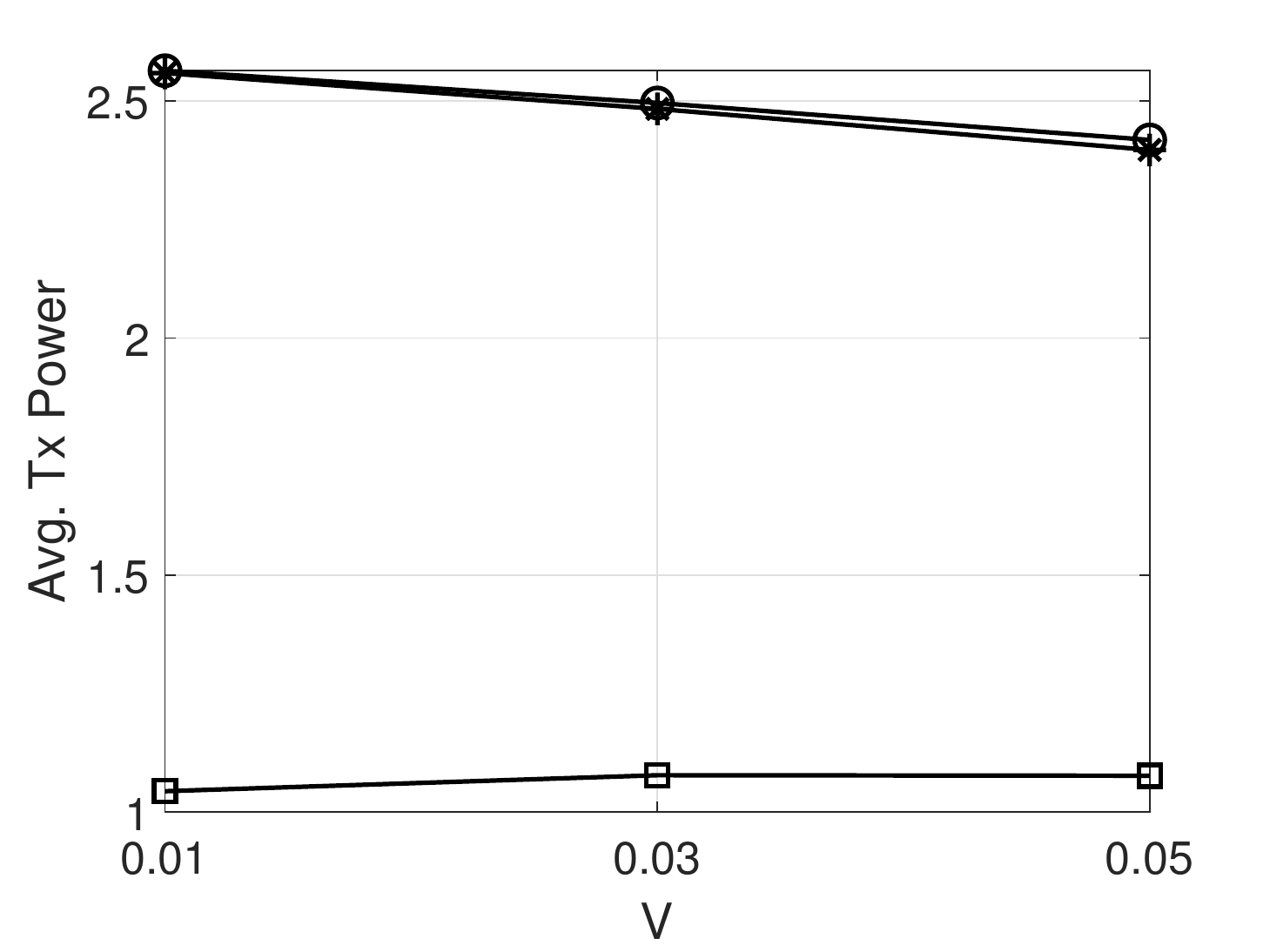} &
	\includegraphics[page=1, width =1\linewidth]{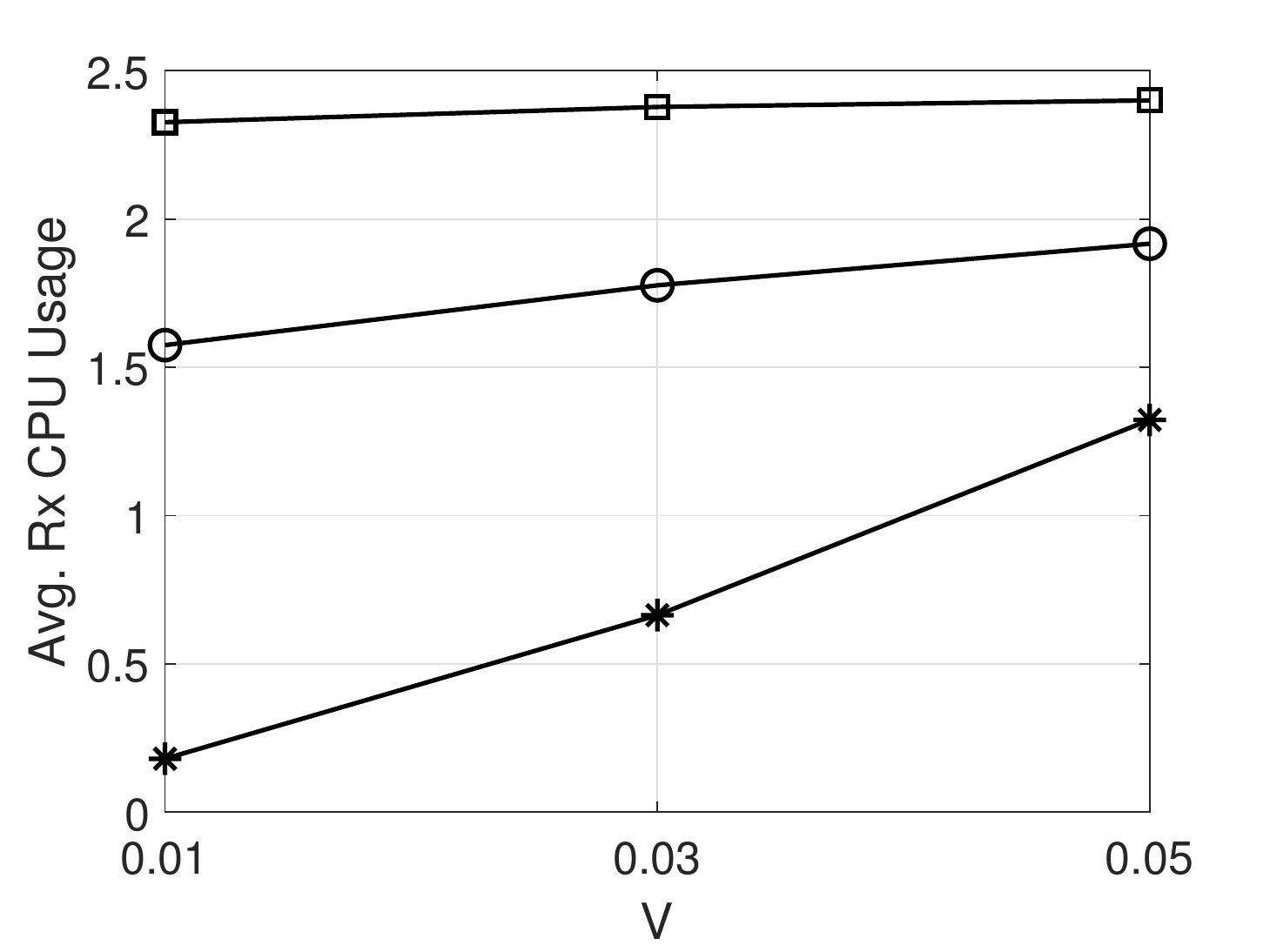} &
	\includegraphics[page=1, width =1\linewidth]{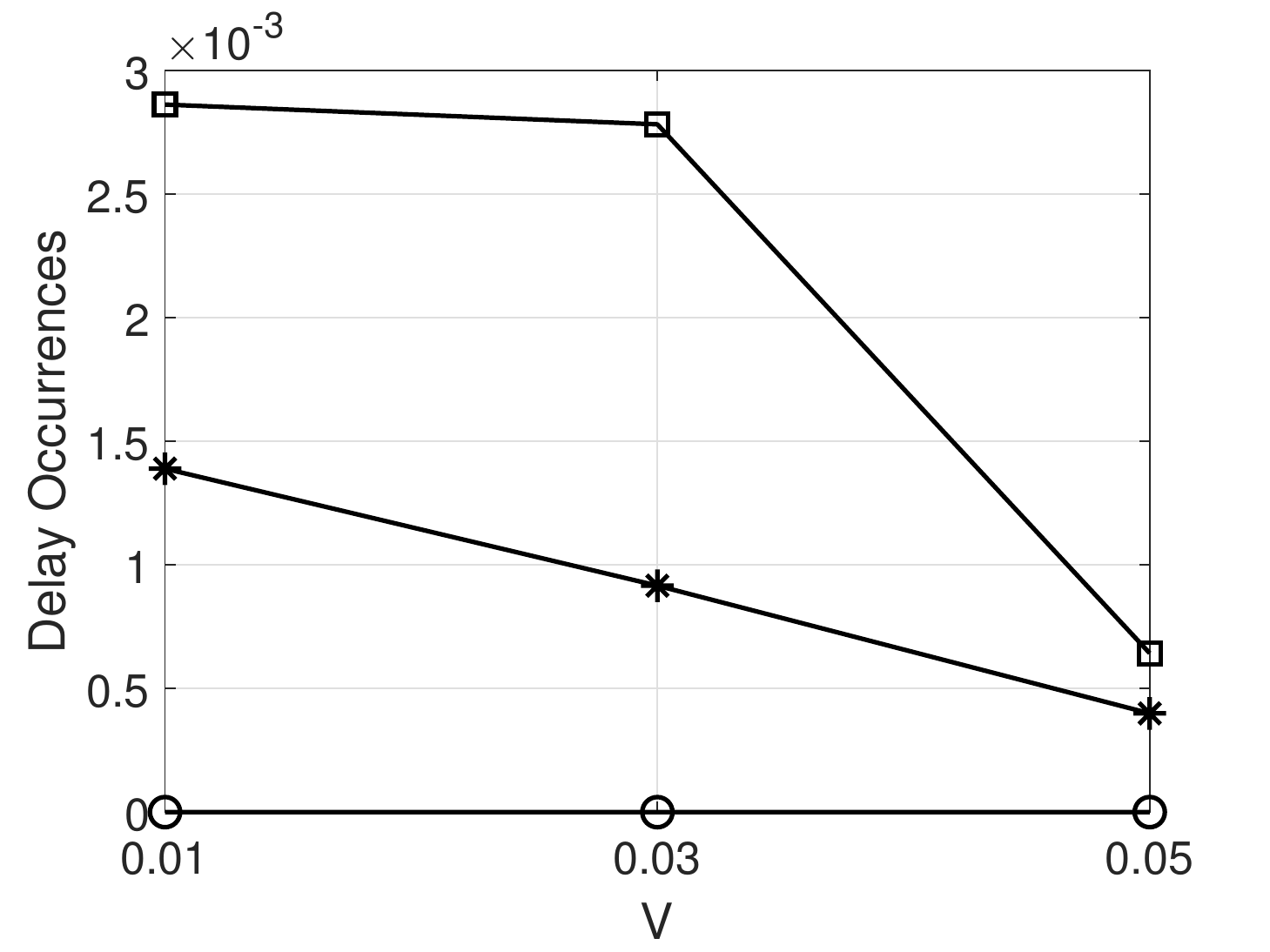}
     \tabularnewline
    \centering (d) Average Tx power consumption &
    \centering (e) Average Rx CPU usage &
    \centering (f) Delay occurrences
\end{tabular}
\caption{Performance of the adaptive quality control of video streaming system vs. $V$.}
\label{fig:result_v}
\end{figure*}

In Fig. \ref{fig:result_v}, we can see how the performance metrics vary with the system parameter $V$, which is a weight factor for the video quality degradation term in the \textit{drift-plus-penalty} term in \eqref{eq:dpp}, i.e., $(\overline{\mathcal{P}} - \mathcal{P}(r(t), d(t))) N(t)$.
In this case, as $V$ grows, the transmitter delivers the small number of chunks which is not transcoded much.
Meanwhile, although the number of transmitting chunks is small, their file sizes would be large because of their small transcoding rates so that the transmit power could not change much. 
At the receiver side, as large $V$ pursues the high-quality streaming, more depths of the ASRGAN and more CPU cores are chosen for enhancing the quality of received chunks.
Even though the receiver operates the deeper ASRGAN when $V$ is large, the transmitted chunks are not compressed much at the transmitter and the receiver also uses more CPU clocks; therefore, the buffer length and the delay incidence would decrease.
Fig. \ref{fig:result_v} shows the above phenomena that occur as $V$ grows.

In summary, the proposed video streaming system improves the average quality while limiting the queuing delay and the streaming delay and utilizing the available power and CPU resources. 
First, compared to `Comp1', the proposed scheme provides almost the same performances of the queue length and transmit power consumption that are related to transmitter decisions (i.e., $N(t)$, $r(t)$ and $P(t)$). Meanwhile, the streaming delay incidence and the average quality measure (i.e., PSNR) of the proposed scheme that are dependent on  both transmitter and receiver decisions are better than those of `Comp1'. 
In addition, since $\xi=2.5$, the proposed scheme utilizes its CPU more efficiently than `Comp1'. 
Here, we can see that jointly controlled decisions of the transmitter and the receiver allow to strike a balance among a variety of conflicting performance metrics carefully.
Second, we show how the adaptive SR could provide the smooth video streaming even with poor channel conditions and/or limited computing resources by comparing with `Comp2'. 
Even though `Comp2' also adjusts its CPU usage under $\xi=2.5$, non-adaptive SR always requires heavy computational tasks resulting in the long buffer length and the frequent incidence of the streaming delay.


\section{Conclusion}
\label{sec:conclusion}

This paper proposes a novel delay-sensitive and power-efficient quality control algorithm for dynamic video streaming using the adaptive SR. 
We first apply the adaptive neural network to SR operation which allows to adaptively choose the depth number of the ASRGAN for enhancing the quality of the received chunks so that computational tasks required for SR can be reduced depending on available computing resources with the expense of the quality of output images.
With the capability of transcoding and adaptive SR operation at the transmitter and receiver sides, respectively, we finally adaptively control the compression rate of transmitted chunks and computational tasks for SR operation depending on the transmitter queue, the receiver buffer, the transmitter power consumption, and the receiver CPU usage. 
Simulation results show that the proposed scheme strikes a balance among a variety of conflicting performance metrics of video streaming carefully, i.e., video quality, queuing delay, chunk processing delay, playback stall, power consumption, and CPU usage.

\ifCLASSOPTIONcompsoc
  \section*{Acknowledgments}
\else
  \section*{Acknowledgment}
\fi
The work was financially supported in part by the Institute for Information \& Communications Technology Promotion Grant funded by the Korea government (MSIT) 2018-0-00170, Virtual Presence in Moving Objects through 5G, and by the National Research Foundation of Korea under Grant NRF-2020R1G1A1101164.

\ifCLASSOPTIONcaptionsoff
  \newpage
\fi

\end{document}